\documentclass[usenatbib]{mn2e}
\usepackage{times}
\usepackage{epsfig}
\usepackage{colordvi}
\usepackage{graphicx}
\usepackage{amssymb,textcomp}
\usepackage{color}

\newcommand{\pasj}{PASJ}
\newcommand{\pasp}{PASP}
\newcommand{\apjs}{ApJS}
\newcommand{\apj}{ApJ}
\newcommand{\apjl}{ApJL}

\newcommand{\aap}{A\&A}
\newcommand{\aj}{AJ}
\newcommand{\mnras}{MNRAS}
\newcommand{\nat}{Nature}

\newcommand{\araa}{ARA\&A}
\newcommand{\aapr}{A\&ARv}          
\newcommand{\aplett}{Ap. Letters}          
\newcommand{\apss}{ApSS}

\newcommand{\msun}{\mbox{M}_{\odot}}       
\newcommand {\beq}{\begin {eqnarray}}
\newcommand {\eeq}{\end {eqnarray}}

\title[Hot accretion flow in black hole binaries]
{Hot accretion flow in black hole  binaries: 
a link  connecting X-rays to the infrared}
\author[Alexandra Veledina, Juri Poutanen \& Indrek Vurm]
    {Alexandra Veledina,$^{1}$\thanks{E-mail: alexandra.veledina@gmail.com, juri.poutanen@oulu.fi} 
    Juri~Poutanen$^{1}$\footnotemark[1] and Indrek Vurm$^{2,3}$ \\
$^1$Astronomy Division, Department of Physics, PO Box 3000, FIN-90014 University of Oulu, Finland \\
$^2$Racah Institute of Physics, Hebrew University of Jerusalem, 91904 Jerusalem, Israel \\
$^3$Tartu Observatory, 61602 T\~{o}ravere, Tartumaa, Estonia}

\begin{document}

\date{Accepted 2013 January 18. Received 2013 January 18; in original form 2012 September 30}
\pagerange{\pageref{firstpage}--\pageref{lastpage}}
\pubyear{2013}
\date{\today}

\maketitle
\label{firstpage} 

\begin{abstract}
\noindent
Multiwavelength observations of Galactic black hole transients have opened a new path to 
understanding the physics of the innermost parts of the accretion flows.
While the processes giving rise to their X-ray continuum have been studied extensively, 
the emission in the optical and  infrared (OIR) energy bands was less investigated and remains poorly  understood. 
The standard accretion disc, which may contribute to the flux at these wavelengths, 
is not capable of explaining a number of observables: 
the infrared excesses, fast OIR variability and a complicated correlation with the X-rays.
It was suggested that these energy bands are dominated by the jet emission, 
however, this scenario does not work in a number of cases.
We propose here an alternative, namely that most of the OIR emission is produced 
by the extended hot accretion flow. 
In this scenario, the OIR bands are dominated by the synchrotron radiation from the non-thermal electrons.
An additional contribution is expected from the outer irradiated part of the accretion disc heated by the X-rays. 
We discuss properties of the model and compare them to the data. 
We show that the hot-flow scenario is consistent with many of the observed spectral data,
at the same time naturally explaining X-ray timing properties, fast OIR variability and its 
correlation with the X-rays.
\end{abstract}


\section{Introduction}

Although the black hole X-ray binaries (BHBs) have  been intensively studied  for over four decades, 
many problems remain unsolved. 
Among the most debated topics are the physics of state transitions, the interplay between the cold accretion disc and the hot
medium, the role of the jet, the source of rapid variability, radiative processes shaping the broadband spectrum and, 
specifically, the nature of various components contributing to its different parts.  
When addressing the latter problem, three distinct components are usually considered: the standard (or irradiated) cool
accretion disc, the hot inner flow (or corona) and the jet.
Their relative contribution depends on the spectral energy range and varies with time and can be assessed by performing
(quasi-) simultaneous multiwavelength observations.

Over the past decade, numerous multiwavelength campaigns have resulted in a significant progress in the field. 
Broadband radio to X-ray spectral energy distributions (SEDs) for many black holes (BH)  
low-mass X-ray binaries (LMXBs) were constructed
\citep[e.g.][]{HMH00,MHG01,CHM03,Cadolle07,CB11,DGS09}. 
In addition to the spectral information, data on the fast variability are now available in the X-rays as well as at lower energies.
The light curves  in the optical/infrared (OIR) and ultraviolet (UV) bands are significantly correlated with the X-rays
\citep{Kanbach01,HHC03,HRP06,HBM09,DGS08,DSG11,GDD10} showing  a complex shape of the cross-correlation function (CCF). 
It provides an important information on the interrelation between various components and gives clues to their physical origin. 

The radio emission in BHBs is likely dominated by the jet as supported by the observed linear polarization at a 1--3 per cent
level in the hard state \citep{Corbel00} and up to ten per cent in spatially resolved components during the transient events
\citep{FGM99,HHC00}.
In addition, a relatively high luminosity, requiring the size exceeding the typical binary separation \citep{Fender06},
as well as the detection of  superluminal motion \citep{MR94} lean towards this interpretation.  
The power-law-like radio spectrum is often attributed to synchrotron emission of an inhomogeneous source in analogy with the
extragalactic jets \citep{BK79}.
In blazars, the jet is also responsible for the X-ray and $\gamma$-ray production \citep*{Konigl81,DS93,SBR94,StP06}. 
On the contrary, the jets in BHBs are unlikely to be responsible for bulk of the X-ray photons 
\citep[for comprehensive discussion see][]{PZ03,ZLG03}. 
Further, the medium producing the X-ray radiation can neither form the base of the jet \citep*{MBF09}, nor be a jet-dominated
accretion flow \citep{Maccarone05}.

The spectra of hard-state BHBs constitute a power-law in the X-ray band with a stable spectral slope and ubiquitous sharp
cut-off at $\sim$100~keV \citep{G97,ZPM98,IPG05}.
It is broadly accepted to be produced by thermal Comptonization \citep[e.g.][]{P98,ZG04}.
Additionally, a Compton reflection feature originating from cool opaque matter (likely the cool accretion disc) is often
detected.
Its strength is correlated with the X-ray slope (\citealt*{ZLS99}; \citealt{ZLG03}), with the width of the iron line as well as
with the quasi-periodic oscillation (QPO) frequency \citep*{GCR99,RGC01,Gilfanov10}. 
These observations support a view that the X-rays are produced in the very vicinity of the BH, 
in a hot flow surrounded by the cold disc. 
In this scenario, variations in the mass accretion rate are correlated with the cool disc truncation radius
\citep*{Esin97,PKR97}, with the flux of soft seed photons that determines the spectral slope and with the reflection amplitude
that scales with the solid angle at which the cold disc is seen from the hot flow.  
Correlations with the QPO  frequency are also naturally explained if the oscillations are produced in the innermost part of
the accretion flow by Lense--Thirring precession \citep{ID11}.
Such a scenario would favour models where seed photons for Comptonization are provided by the standard \citet{SS73} accretion
disc.
However, the hot flow itself also produces synchrotron radiation that can contribute or even dominate the seed photon flux to
the Comptonizing medium (\citealt*{GHS98}; \citealt{WZ00,WZ01,MB09,PV09,SPDM11}; \citealt*{VVP11}).  

Discovery of the high-energy (MeV) tails in the hard-state accreting BHBs 
(\citealt{McConnell94,McConnell02,Ling97,DBM10}; \citealt*{JRM12}) 
suggests the presence of non-thermal particles in these systems.
Such particles may be produced in a hot inner flow or a jet.
Their association with the jet, however, is inconsistent with detections of even more prominent high-energy non-thermal tails
in the soft state of BHBs \citep{Grove98,Gier99,ZGP01,McConnell02,GD03}, when the jet is quenched \citep*{FBG04}.  
In this state the inner flow is likely to be replaced by a corona, which remains here the only alternative. 
In the hard state, the entire X/$\gamma$-ray spectra can be produced by hybrid (thermal plus non-thermal) electrons
via synchrotron self-Compton (SSC) mechanism \citep{MB09,PV09}.  
The thermal part of the particle distribution is responsible for the power-law-like Comptonization 
continuum with the sharp cut-off, while the
non-thermal particles both produce seed synchrotron photons for Comptonization and contribute to the MeV energies via inverse
Compton process.
Transition to the soft state can then be associated with the rising role of the disc as a source of seed photons, which
increases Compton cooling and causes changes in the electron distribution from mostly thermal to nearly non-thermal
\citep{PC98,PV09,VVP11}. 
The SSC mechanism was also shown to be consistent with the peculiar optical variability, which in a number of BHBs  
is partially anticorrelated with the X-ray emission \citep{Kanbach01,DGS08,GMD08}. 
Namely, the increasing mass accretion rate results in a higher X-ray and a lower synchrotron OIR emission, because 
of an increasing role of synchrotron self-absorption within the source \citep*{VPV11}.  

The OIR spectra of LMXBs often show an excess above the standard accretion disc (e.g., \citealt{HMH00,HHC02}; \citealt*{GGH10}).
In some cases the spectrum can be described by a power-law of index close to zero (i.e. $F_\nu\propto \nu^0$).
Such data were previously explained by additional contribution from the irradiated disc \citep*{GDP09}, 
dust heated by the secondary star \citep{MM06} or the jet \citep{HHC02,GMM07}.
We show that they also can be explained by the synchrotron radiation from the non-thermal particles in the hot flow. 
However, in some cases the OIR fluxes are higher than expected from any candidate alone \citep{CHM03,GDD10}, 
suggesting contribution of at least two components simultaneously. 
This can be a reason for the complex shape of the optical/X-ray CCF \citep{VPV11}.

The general shape of the time-averaged  X/$\gamma$-ray spectrum of BHB can be well explained in terms 
of a one-zone hybrid Comptonization model.  
However, the short-term spectral variability, reflected in hard X-ray time-lags \citep{MiKi89,Nowak99a}, and asymmetries of the
CCF between hard and soft X-ray energy bands (\citealt{PGR79,NGM81}; \citealt*{MCP00}) suggest that a
number of regions simultaneously contribute to the total spectrum.
The observed logarithmic dependence of the time-lags on photon energy can be phenomenologically explained by spectral pivoting
\citep{PF99}.
Theoretical model capable of explaining the observed timing properties was proposed by \citet*{KCG01} and further developed
in \citet{AU06}.
It assumes that the X-ray spectrum is produced in the hot flow/corona, present in a range of radii, by Comptonization of the disc
photons. 
The power-law slope of the locally emitted spectrum depends on the distance from the BH: the hardest spectra are produced
in the innermost region due to the lack of photons from the cold disc.
The main source of the short-term variability in BHs is believed to be fluctuations in the mass accretion rate, 
propagating through the hot accretion flow \citep{Lyub97}.
The hard time-lags thus naturally appear from the perturbations propagating from the larger distances to the 
vicinity of the BH. 
The model of \citet{KCG01} considers thermal Comptonization of the disc photons, but as we show below the results hold in the
framework of hybrid Comptonization, with the synchrotron mechanism as the major seed photon supplier. 

It is clear that the complete description of the timing and spectral properties of BHBs requires a multizone model.
In this paper we construct such a model for the extended hot accretion flow. 
This model is somewhat analogous to the inhomogeneous synchrotron models developed for extragalactic jets
\citep{Marscher77,BK79}.
The difference is that in our model the emission originates from an inflow, not an outflow.
The advantage of the hot-flow model is that the energy input can be estimated from the available gravitational energy
transferred to particles via some mechanism, while in the jet scenario the energy release is an arbitrary function.

Similarly to the previously studied one-zone models \citep{MB09,PV09,VVP11}, 
we assume that gravitational energy is dissipated
in the flow and injected in the form of electrons having a power-law  distribution 
(while the steady-state electron distribution is mostly thermal).
We compute the steady-state particle and photon distributions self-consistently by solving corresponding kinetic equations. 
Our aim is to understand the broadband spectral properties of such a flow and compare them to the BHB data.

\begin{figure*}
\centerline{ \epsfig{file=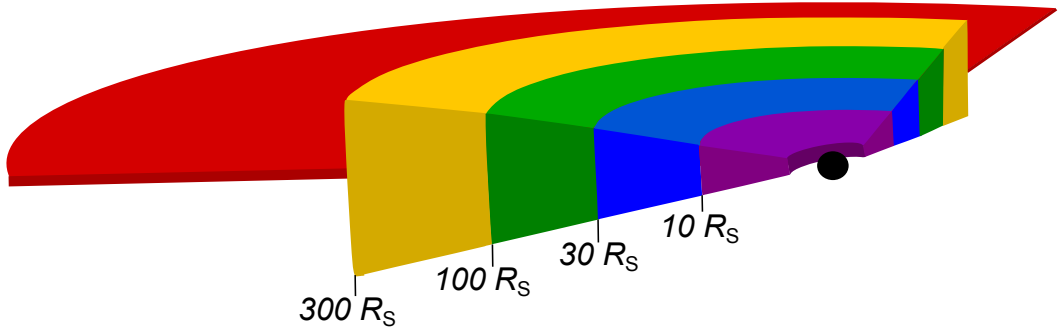, width=14cm}}
\caption{Schematic picture of the accretion flow inner regions.
Red outer component represents the multi-color cold accretion disc, truncated at 300 Schwarzschild radii ($R_{\rm S}$).
The inner parts are occupied by a geometrically thick hot accretion flow. 
In our numerical model,  we split the inner hot flow into four zones with outer radii 
$10R_{\rm S}$ (violet), $30R_{\rm S}$ (blue), $100R_{\rm S}$ (green) and $300R_{\rm S}$ (yellow).
}
\label{fig_inner_flow}
\end{figure*}

In Section~\ref{sect:model}, we give constraints on the size of the hot accretion flow that can be derived from the observed
level of the OIR emission. 
We first construct an analytical model for the hot flow assuming power-law radial dependences of the main parameters. 
We then proceed to the numerical model where the electron distributions 
and the emitted spectra are computed self-consistently.
In Section~\ref{sect:synch_num}, we present the results of simulations for the model corresponding to the hard state of BHBs.
We show that the multizone hot disc model produces flat OIR  spectra resulting from synchrotron emission of non-thermal
electrons at different radii.
We then model  the state transitions by decreasing the truncation radius of the hot flow. 
In Section~\ref{sect:comparison}, we provide a detailed analysis of the observational data and compare them to our model and to
the jet scenario.
We summarize our finding in Section~\ref{sect:summary}.

\section{Analytical model}
\label{sect:model}

\subsection{Geometry}

Many observational properties suggest that the standard disc in the hard state is truncated far away from the central object 
\citep[for detailed description and challenges to the truncated disc scenario, see review by][]{DGK07}.
The inner part is probably occupied by some type of geometrically thick, optically thin, hot accretion flow, which is
responsible for the X-ray Comptonization continuum, but also contributes to the longer wavelengths. 

One can roughly estimate the minimum size of the source that is required to produce the observed OIR luminosity. 
Let us first assume that OIR emission is produced by synchrotron radiation from thermal particles 
(as for example was discussed earlier by \citealt{FGM82}).
The typical temperature of the electron gas determined from the Comptonization cut-off is $kT_{\rm e}$$\sim$100~keV 
and the typical IR luminosity at $1$~eV can reach $\nu L_{\nu,\rm 1\ eV}=10^{36}$~erg~s$^{-1}$.
Thus, we get the minimum size from the Rayleigh-Jeans formula 
\begin{equation} \label{eq:radius_therm}
 R = \left( \frac {\nu L_{\nu}}{8 \pi^2 \nu (\nu/c)^2 k T_{\rm e}} \right)^{1/2} 
 \approx 2.3 \times 10^{9} \ \mbox{cm}.
\end{equation}
For a 10$\msun$ BH, assumed in all calculations hereafter, this corresponds to 750~$R_{\rm S}$
(here $R_{\rm S}=2GM/c^2$ is the Schwarzschild radius).
However, many observed properties [e.g. iron line width, amplitude of Compton reflection, drop of the iron line equivalent
width with the Fourier frequency, see \citet{Gilfanov10}, as well as dependence of the X-ray time-lags on
energy below $\sim$1~keV, \citet{UWC11}] suggest that the cold disc in the hard state is truncated at a smaller radius.
Moreover, in order to produce sufficient amount of seed photons for Comptonization, an extremely high magnetic field is required 
in addition to the large source size \citep*[see Appendix~\ref{apndx:nth_frac} and also, e.g.,][]{DiMCF99,MDiMF00}.
Thus, the thermal radiation of the hot flow is unlikely to be a good candidate to produce enough OIR photons \citep[see also][]{WZ00}.

However, even a weak, energetically unimportant non-thermal tail, in addition to the mostly Maxwellian distribution,
gives a significant rise to the synchrotron luminosity \citep[e.g.][]{WZ01}.
For example, a tail containing only one per cent of total particle energy increases it by a factor of $100$. 
Accurate calculations (see below) show that the source size of $R\approx(30-50)R_{\rm S}$, an order magnitude smaller than given
by equation (\ref{eq:radius_therm}), would in principle be enough to radiate the observed OIR luminosity. 
Such a size is consistent with the above estimates of the truncation radius and with the typical size of the region of the
gravitational energy release.
It is worth noticing that a strong synchrotron emission from non-thermal particles makes it a good candidate for seed photons
for Comptonization \citep{MB09,PV09}, which implies that the SSC spectrum extends from the X-rays down to the OIR band with
the low-energy turnover determined by the maximum extent of the hot flow.

\subsection{Hard-state OIR spectra} 
\label{sect:synch_th}

Spectral properties of the hot flow in the OIR band can be understood from simple analytical considerations. 
Let us consider the flow with a constant height-to-radius ratio $H/R$, extending between radii $R_{\rm in}$ and
$R_{\rm out}$ (see Fig.~\ref{fig_inner_flow} for the geometry). 
In order to estimate the synchrotron luminosity and spectra, we assume that the electrons follow a power-law distribution in
Lorentz factor $n_{\rm e}(\gamma)\equiv d n_{\rm e}/d\gamma=n_0 \gamma^{-p}$, starting from $\gamma_{\rm le} = 1$ 
to infinity.
Deviations from the power-law at low energies do not play any role, as the synchrotron emission produced
by these electrons is self-absorbed. 
The Thomson optical depth across the disc is assumed to follow the power law $\tau(R)\propto R^{-\theta}$. 
For the constant $H/R$ this is equivalent to $n_0(R)\propto \tau /R \propto R^{-\theta-1}$. 
We further assume that the magnetic field depends on the distance from the BH as  $B(R)\propto R^{-\beta}$. 

Our analytical model of the hot flow is analogous to the non-uniform synchrotron source models, previously applied to the
emission from extragalactic jets \citep*{CD73,deBryun76,Marscher77,BK79,Konigl81,GMT85}.
For the parameters considered here,  most of the luminosity is produced in the inner part of the source, 
so that the OIR spectrum is composed of emission components coming from different radii and is not dominated by the radiation
from the outer regions (these two cases are illustrated in figs 2a and 2b in \citealt{GMT85}).
 
A region of the disc at a given radius emits  synchrotron radiation, which is self-absorbed below the turnover frequency
$\nu_{\rm t}$. 
For power-law electrons, this frequency can be calculated as \citep{RL79,WZ01}
\begin{equation}
\nu_{\rm t} = 3^{\frac{p+1}{p+4}} 2^{-\frac{6}{p+4}} \pi^{\frac1{p+4}} \nu_{\rm L}^{\frac{p+2}{p+4}}
                   \left[ G_1(p) c R r_{\rm e} n_0 \right]^{\frac{2}{p+4}},
\end{equation}
where $\nu_{\rm L}=eB/(2\pi m_{\rm e}c)$ is the Larmor frequency, $G_1(p)\simeq1$ is a combination of Euler's Gamma functions
(due to averaging over electron pitch angles), $r_{\rm e}$ is the classical electron radius.
Substituting the constants, we get 
\begin{equation}\label{eq:turn_over_exact}
 \nu_{\rm t} \approx 
  3\times 10^{15}  B_6^{\frac{p+2}{p+4}}  \left( \sigma_{\rm T} n_0 R \right)^{\frac{2}{p+4}} {\rm Hz},
\end{equation}
where $Q=10^{x}Q_x$ in cgs units.
The term in brackets can also be written as 
$\sigma_{\rm T} n_0 R=\tau(\gamma_{\rm t}) \gamma_{\rm t}^{p}$, with $\gamma_{\rm t}$ being the Lorentz factor of the electrons
emitting at the turnover frequency 
and $\tau(\gamma_{\rm t})=\sigma_{\rm T} R n_{\rm e}(\gamma_{\rm t})$ 
being the  Thomson optical depth per unit Lorentz factor at $\gamma_{\rm t}$.
In the later representation, the equation is also valid for hybrid electrons (e.g., Maxwellian with power-law tail), as long as
the electrons emitting at the turnover frequency are in the power-law tail.
The low-frequency cut-off for synchrotron spectrum from power-law electrons scales as
\begin{equation}\label{eq:turn_over_sca}
 \nu_{\rm t}\propto R^{-[\beta(p+2)+2\theta]/(p+4)}. 
\end{equation}
Again, for hybrid electron distribution one should consider scaling with radius of the power-law tail (parameter $\theta$),
which can be different from scaling of the total optical depth.
As immediately follows from equation (\ref{eq:turn_over_exact}), the turnover frequency may fall to optical and even IR
wavelengths for sufficiently low magnetic field and/or Thomson optical depth.

\begin{figure}
\centerline{ \epsfig{file=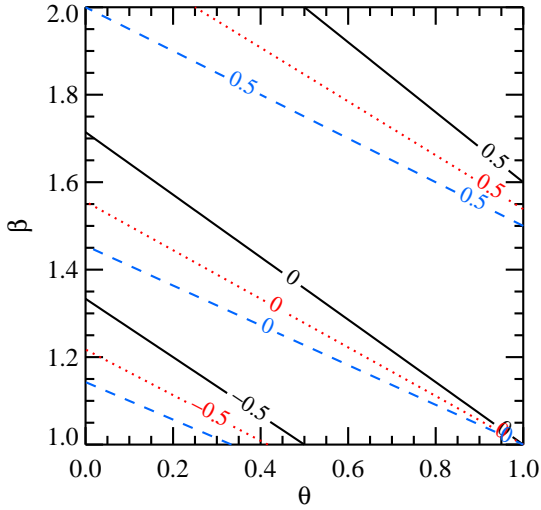, width=7cm}}
\caption{Contour plot of the constant index  $\alpha_{\rm OIR}$ 
as a function of parameters $\beta$ and $\theta$ for 
three values of electron index $p$=$2$ (solid  black), $3$ (dotted red) and $4$ (dashed blue). 
}
\label{fig_alpha_oir}
\end{figure}

The emission at the turnover frequency is optically thick, so the intensity is equal to 
the source function for the power-law electrons.
For isotropic electrons the intensity is (averaged over pitch angles)
\begin{equation}
 I_{\nu_{\rm t}} = \frac{m_{\rm e}G_2(p)}{2\sqrt{3}} \nu_{\rm L}^{-1/2} \nu_{\rm t}^{5/2},
\end{equation}
where $G_2(p)\simeq1$ (again, coming from the angle-averaging). 
At each wavelength, there is a contribution from the optically thick and optically thin emission from different radii.
For simplicity, we assume
that emission from each radius contributes only to its own turnover frequency,\footnote{Precise calculations of additional
contribution from optically thin parts result in a slightly different normalization, while the spectral slope remains the same
\citep[see][]{Marscher77}.} 
therefore the resulting spectrum of an inhomogeneous synchrotron source constitutes a power-law
\begin{equation} \label{eq:luminosity}
 \nu L_{\nu}  = 4\pi^2 R^2 \nu I_{\nu} \approx 
           2\times10^{36}   R_8^2 \  B_6^{-1/2} \   \nu_{15}  ^{7/2}  \;{\rm erg\,s}^{-1}. 
\end{equation}
Substituting the appropriate parameter scaling and using equation~(\ref{eq:turn_over_sca}), we get the spectral index
\begin{equation}\label{eq:OIR_slope}
 \alpha_{\rm OIR}    =    \frac {5 \theta + \beta (2p + 3) - 2p - 8}{\beta (p+2) + 2\theta},
\end{equation}
where $\displaystyle L_{\nu} \propto \nu^{\alpha}$. 
In a wide range of parameters $\beta\in [1,2]$ and $\theta\in [0,1]$ 
the resulting spectral  slope lies between $-0.5$ and $0.5$ (see Fig.~\ref{fig_alpha_oir}).

\section{Numerical model}
\label{sect:synch_num}

The analytical model developed in Section~\ref{sect:synch_th} describes only the OIR \textit{synchrotron} spectra and is valid for
purely power-law electrons. 
Such distributions may result from various acceleration mechanisms e.g. shock acceleration or magnetic reconnection. 
In the limit of low optical depth and weak magnetic field the electrons are unable to cool and the shape of the distribution
stays unchanged. 
These conditions might be satisfied in quiescent state, for which the analytical model can be applied.
During the accretion outbursts, the matter density in the hot flow increases and the energy exchange and cooling processes
become important; thus, the initial power-law distribution evolves.
The most important mechanisms operating in the hot rarefied plasmas of the hot accretion flows are Compton scattering,
synchrotron emission and absorption, Coulomb collisions, bremsstrahlung, and possibly photon-photon pair production and
annihilation. 
At high energies, for a continuously operating acceleration, the steady state distribution remains a power-law-like, 
but softens because of cooling by Compton, synchrotron and bremsstrahlung.
At lower energies, Coulomb collisions and synchrotron self-absorption efficiently thermalize particles, forming a Maxwellian
distribution.
The total particle distribution  consists of a low-energy Maxwellian plus a high-energy tail.
Such a distribution we call hybrid.
The shape and energy content of the tail are fully determined by the balance between acceleration and cooling processes.
It cannot be calculated analytically; therefore, we treat this problem numerically.
The photon spectrum emitted by the hot flow is computed self-consistently with the particle distributions.  

The time-scale of equilibration of electron and photon distributions for typical parameters of our model 
is smaller than the corresponding advection time in the hot flow (see Appendix~\ref{apndx:times}).
Thus we can use an assumption that the electron and photon distributions  are in a steady state. 
We obtain them by solving  the relevant kinetic equations.

\begin{figure}
\centerline{ \epsfig{file=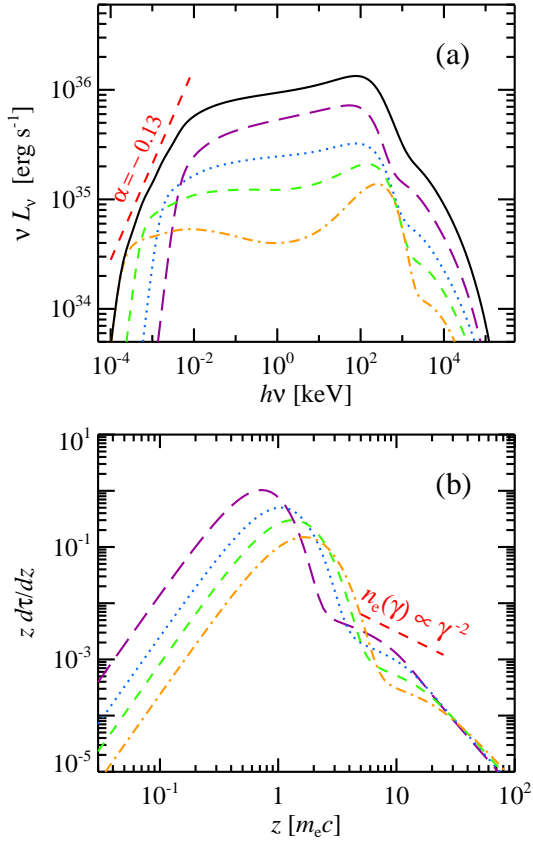, width=7cm}}
\caption{
Photon spectra (upper panel) and electron distributions (lower panel) for the hard-state model 
with initial electron injection index $\Gamma_{\rm inj}=2.5$.
Electron momenta $z=\sqrt{\gamma^2-1}$ are measured in units of $m_{\rm e}c$.
Other parameters are listed in  Table~\ref{tab_params}.
The lines correspond to zone~1 (long-dashed), zone~2 (dotted), zone~3 (dashed) and zone~4 (dot-dashed).
Sum of the components is shown with a solid line.
The red short-dashed line shows the slopes from analytical approximation.
For further details, see Section~\ref{sect:synch_num_hard}.
}
\label{fig_pureSSC_inj2_5}
\end{figure}

\begin{figure}
\centerline{ \epsfig{file=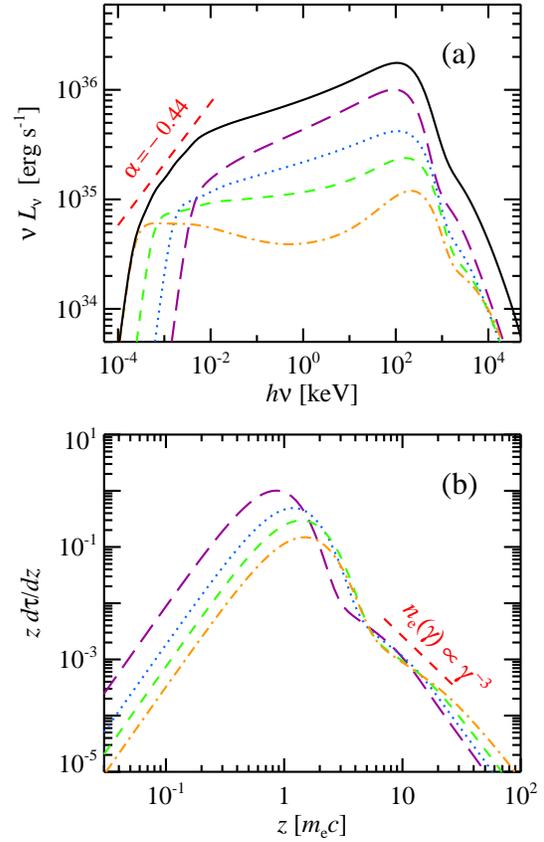, width=7cm}}
\caption{
Same as Fig.~\ref{fig_pureSSC_inj2_5}, but for the electron injection index $\Gamma_{\rm inj}=3.0$.
X-ray spectra are harder in this case, while the OIR spectrum is softer.
}
\label{fig_pureSSC_inj3}
\end{figure}

\subsection{Model set up}
\label{sect:setup}

We consider a geometrically thick optically thin inner accretion flow, which vertical extent is parametrized by the ratio
$H/R=\rm const$.
We assume that the hot flow corresponds to some type of radiatively inefficient accretion flow 
\citep*[see review in ][]{KFM08}. 
In such  flow, the radiative loss rate per unit area scales with radius as $\displaystyle Q_{\rm rad}\propto R^{-5/2}$, and the
electron number density scales as $\displaystyle n_{\rm e}(R)\propto R^{-3/2}$. 
The Thomson optical depth along the vertical direction  $\tau = \sigma_{\rm T} n_{\rm e}(R) H \propto {R}^{-1/2}$. 
The scaling of the magnetic field with radius is model-dependent \citep[e.g.,][]{SK05,Meier05,AF06}.
Here we assume that the magnetic pressure and the radiation pressure are equal throughout the flow, from which we get 
$B\propto R^{-5/4}$.
The latter scaling is the same as in the accretion flow model of \citet{SK05}.
The hot flow extends from $3R_{\rm S}$ to the truncation radius $R_{\rm tr}$, where the cold disc with luminosity $L_{\rm disc}$ 
and colour  temperature $T_{\rm col}$ starts.

The energy transfer to electrons is simulated as a power-law injection with slope 
$\Gamma_{\rm inj}$, extending between the Lorentz factors 1 and $10^3$.\footnote{By using such approximation, we
implicitly assume that 100 per cent of the dissipated energy is transported to particles by acceleration processes, while in
reality most of the energy is likely to be given to particles in terms of heating by diffusive processes, e.g. Coulomb collisions with protons.
In Appendix~\ref{apndx:nth_frac}, we discuss the validity of such an approximation and show that the results hold 
due to efficient electron thermalisation at low Lorentz factors.} 
The model  has seven parameters: (i) the total luminosity $L$, 
(ii) the index of the electron power-law injection spectrum $\Gamma_{\rm inj}$ 
(constant throughout the flow), (iii) the electron Thomson optical depth $\tau_1$ and 
(iv) the magnetic field $B_1$ in the innermost regions, 
(v-vi) indices of their power-law radial dependencies $\theta$ and $\beta$, and 
(vii) the hot-flow size $R_{\rm tr}$.

The energy given to electrons is redistributed between the particles (electrons and positrons) and 
photons in processes of synchrotron
emission and self-absorption, Compton scattering, Coulomb collisions, pair production and annihilation 
and bremsstrahlung emission. 
The dominating cooling regime for a specific electron Lorentz factor depends on the
luminosity, magnetic field and the optical depth (relevant scaling is given in Appendix~\ref{apndx:times}).
In addition to the internally produced radiation, we also consider soft photons from the 
cold outer accretion disc in the form of the blackbody radiation injected homogeneously into the system.
The kinetic equations for electrons and photons describing relevant radiative processes are solved 
using the code developed by \citet{VP09}. 

To compute the radiative transfer in the , we divide it 
into a number of separate regions/zones (Fig.~\ref{fig_inner_flow}).
Each zone $i$ has size (in the radial direction) $\Delta R_{i} = R_{i+1}-R_{i}$ equal to its full height 
at the zone centre $2H_{i}=  2 R_{i, \rm c} (H/R)$ 
(where $R_{i, \rm c}=(R_{i}+R_{i+1})/2$ is the distance to the center of $i$th zone), implying 
\begin{equation}\label{eq:Rii1}
R_{i+1} = R_{i}\frac{1+H/R}{1-H/R}.
\end{equation}
The  net energy input into the $i$th zone equals to its luminosity: 
\begin{equation}
  L_i  = 2\pi \int \limits_{R_{i}}^{R_{i+1}} Q_{\rm rad}(R) R dR
     \propto \frac{1}{\sqrt{R_{i}}} - \frac{1}{\sqrt{R_{i+1}}}. 
\end{equation}
The characteristic Thomson optical depth of the $i$th zone is associated with  that in 
the vertical extension 
\begin{equation}
 \tau_i = \sigma_{\rm T} n_{\rm e}(R_{i,\rm c}) H_i. 
\end{equation}

Additional soft photons from the outer cold disc 
are described by the colour temperature $T_{\rm col}$ and the disc luminosity coming to the $i$th zone
$(\Omega_{i}/4\pi)L_{\rm disc}$. 
Here the factor $\Omega_{i}$ accounts for the fact that only a part of the disc luminosity is entering the hot flow.
It is fully determined by the cold disc/hot-flow geometry and in our case can be approximated as 
\begin{equation}
\frac{\Omega_{i}}{4\pi}\approx \frac {1}{4} \left( \frac{R_{i+1}}{R_{\rm tr}} \right)^3,
\end{equation}
where 
$\pi (R_{i+1}/R_{\rm tr})^2$ is the solid angle of the $i$th zone as
seen from the cold accretion disc, and another factor of $R_{i+1}/R_{\rm tr}$  
accounts for anisotropy of the disc radiation. 
This formula is accurate for the  zone adjacent to the disc. 
It overestimates the contribution of disc photons to the innermost zones, 
but in that case $\Omega_i$ is very small and the disc contribution is negligible.

Each zone represents a torus-like structure with the major radius $R_{i,\rm c}$ and the minor radius $H_i$. 
The radiative transfer is handled under the local approximation of homogeneous isotropic distributions in a sphere with 
radius $H_i$, using escape probability method \citep[see][]{VP09}.  
The power injected  into the sphere is scaled proportionally to the ratio of 
respective  volumes of the sphere and the torus: 
\begin{equation}
 L_{i,\rm sph} = \frac{V_{i,\rm sph}}{V_i} L_i = \frac{2}{3\pi} \frac{H}{R} L_i,
\end{equation}
where $V_{i,\rm sph}= \frac{4\pi}{3}H_i^3$ is the sphere volume, $V_i$ and $L_i$ are volume and luminosity of the $i$th zone.
This approach keeps the energy density inside the sphere and the torus the same. 
After the spectrum in a sphere is computed, we multiply it by the same factor $V_i/V_{i,\rm sph}$ in order to account for the radiation
from entire torus.
The total spectrum of the flow is the sum of the spectra from each zone. 

This local approach neglects the interaction between different zones. 
The influence of the outer zones on the inner zone spectra is negligible because 
of their lower  luminosity as well as very small solid angle occupied by the inner zones as seen from the outside. 
Although the  effect of the inner zones on  the outer zone spectra is more significant, 
the overall spectral properties are practically the same because 
the X-ray spectrum is dominated by the inner zone and the OIR spectral shape is 
determined by the parameter scaling rather than their precise values.   
The radiative transfer effects are considered in Appendix \ref{apndx:transfer}.

\subsection{Hard state}
\label{sect:synch_num_hard}

In the hard state, the hot flow can extend to large radii $\gtrsim 100R_{\rm S}$ and the 
role of the soft photons from outer cold accretion disc is negligible. 
Therefore, we neglect them in the simulations and consider only the emission from the hot flow. 
We take the total luminosity of the flow $L=10^{-2}L_{\rm Edd}$ ($L_{\rm Edd}$ is the Eddington
luminosity), the Thomson optical depth of the innermost zone $i=1$ $\tau=1.25$, typically 
found from the X-ray/$\gamma$-ray data \citep[e.g.][]{ZPM98,FZA01},  magnetic field in the innermost zone 
$B=10^6$~G, and the height-to-radius ratio $H/R=0.5$. 
The radial dependencies of the parameters are given in Sect.~\ref{sect:setup} 
and listed  for each zone  in Table~\ref{tab_params} (first five rows).
The results of simulations are shown in Fig.~\ref{fig_pureSSC_inj2_5} 
for the injection slope $\Gamma_{\rm inj}=2.5$ and in Fig.~\ref{fig_pureSSC_inj3} for $\Gamma_{\rm inj}=3.0$.

\begin{table}
\centering
\caption{Parameters of the multizone hot inner flow model.}\label{tab_params}
    \begin{minipage}[c]{6.0cm}
\begin{tabular}{@{}lcccc@{}}
\hline
Parameter / zone                & 1          & 2       & 3          & 4  \\
\hline
\smallskip
$R_{i}/R_{\rm S}$               & 3          & 10      & 30         & 100 \\
$R_{i+1}/R_{\rm S}$             & 10         & 30      & 100        & 300 \\
$\tau_{i}$                      & 1.25       & 0.65    & 0.4        & 0.2 \\
$B_{i}$ (10$^6$ G)              & 1          & 0.25    & 0.06       & 0.015 \\
$L_i$ (10$^{36}$\ erg s$^{-1}$) & 6          & 3       & 2          &  1 \\
$kT_{\rm col}$$^a$ (keV)        & 0.25       & 0.12    & 0.05       & -- \\
\hline
\end{tabular}
 \begin{flushleft}{
$^{a}$  $kT_{\rm col}$ is the colour temperature of radiation coming into the hot flow
from the inner edge of the cold accretion disc extending down to $R_{i+1}$.
}\end{flushleft} 
\end{minipage}
\end{table}

Simulations show that larger zones generally have softer spectra, 
with difference in the X-ray spectral indices $\Delta \alpha_{\rm X} \approx 0.05 - 0.08$.
The main reason is that the outer zones are more transparent to the synchrotron radiation, 
which increases the ratio of the synchrotron to the thermal Compton luminosities. 
At the same time we see that the equilibrium electron temperature grows with radius 
from approximately 70~keV  up to 240~keV (Fig.~\ref{fig_pureSSC_inj2_5}). 
This is caused by a significant drop of the optical depth $\tau$ in the outer zones, 
with a relatively slow change of the Compton $y$-parameter, defined as $y=4(kT_{\rm e}/m_{\rm e}c^2)\tau$.  
The X-ray spectrum of the outer zones is dominated by thermal bremsstrahlung, 
because its role relative to Compton cooling $\propto \tau R/L$ grows linearly with radius \citep[eq. (22) in ][]{VVP11}.
The combined spectrum of all zones has a concave shape,  exactly as  observed \citep{IPG05}.

The high-energy tail above a few $100$~keV is dominated by Comptonization 
produced by the non-thermal electron tail. 
At Lorentz factor above $20$, the tail has a power-law shape 
corresponding to index $p=\Gamma_{\rm inj}+1$ due to synchrotron and Compton cooling. 
At intermediate $\gamma$, the distribution is curved, because of the large 
role of Coulomb collisions  which produce equilibrium distribution with index $p=\Gamma_{\rm inj}-1$ 
\citep[eq. 12 in ][]{VVP11}.

The OIR spectrum is produced by a combination of synchrotron self-absorption peaks from different zones.
The outer zones dominate at longer wavelengths. 
The low-energy cut-off is determined by the size of the largest zone (equation~\ref{eq:turn_over_exact}), which for
$R=300R_{\rm S}$ and considered values of the parameters (in particular, the assumed magnetic field in the outermost zone) is at
0.2~eV.  At even lower energies, the spectrum is $L_\nu\propto \nu^{5/2}$. 
Above $10$~eV, emission from all zones is optically thin and is dominated by thermal Comptonization of seed non-thermal
synchrotron photons.

The X-ray spectra are harder for $\Gamma_{\rm inj}=3.0$. 
This is a direct consequence of the softer equilibrium electron distribution, which results in a lower synchrotron luminosity
and larger Compton $y$-parameter. 
The OIR spectra are, however, softer in this case, because a different slope of the electron non-thermal tail results in a
relatively low normalization of the electron distribution and, respectively, in the weaker synchrotron emission from the inner zones.

Many of the numerical results can be understood from the analytical model if one approximates the electron distribution by a
power-law  in the energy range where electrons emit close to the self-absorption frequency.  
In our simulations these electrons have Lorentz factors $\gamma_{\rm t}\approx10$.

In the case of injection slope $\Gamma_{\rm inj}=2.5$ the optical depth of the power-law electrons scales the same way as the
total optical depth, with index $\theta=1/2$ (see Fig.~\ref{fig_pureSSC_inj2_5}b).
The slope of the electron distribution is approximately $p=2$.
Putting these parameters (with $\beta=5/4$) into equation~(\ref{eq:OIR_slope}), we get $\alpha_{\rm OIR}=-0.13$, in good
agreement with the numerically computed slope (see Fig.~\ref{fig_pureSSC_inj2_5}a).

For softer electron injection $\Gamma_{\rm inj}=3.0$, we find that the optical depth at the Lorentz factor 
$\gamma_{\rm t}=10$ is nearly constant for every zone (see Fig.~\ref{fig_pureSSC_inj3}b), thus for analytical
approximation we take $\theta=0$.
The average electron slope at this Lorentz factor is $p\approx 3$.
Putting these coefficients into equation~(\ref{eq:OIR_slope}), we get $\alpha=-0.44$, also in good agreement with the computed
spectrum (Fig.~\ref{fig_pureSSC_inj3}a).
 
\begin{figure*}
\epsfig{file=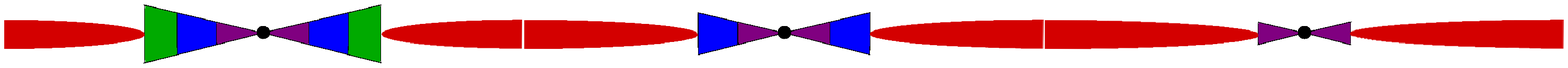, width=16.5cm}\\
\epsfig{file=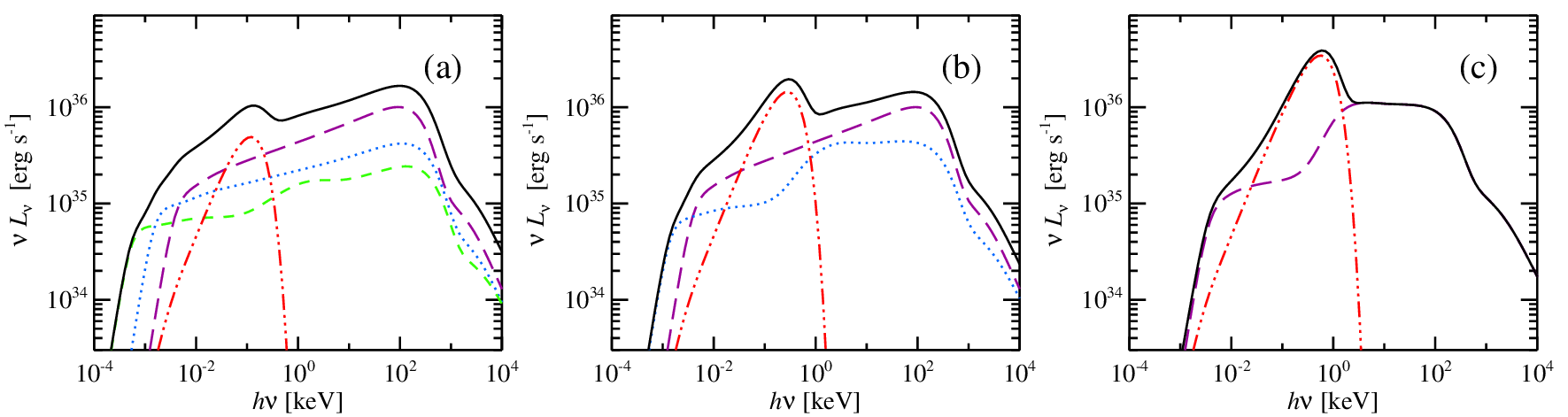, width=16.5cm}
\caption{
\textit{Upper panels}: geometrical evolution at state transition. Principal components are standard accretion disc (red) and
inner hot flow: zone~1 (within $10R_{\rm S}$, violet), zone~2 (within $30R_{\rm S}$, blue) and zone~3 (within $100R_{\rm S}$,
green).
\textit{Lower panels}: spectral evolution at state transition.
Contribution of different zones are marked with lines: zone~1 (long-dashed), zone~2 (dotted), zone~3 (short-dashed) and
thin accretion disc (three-dot-dashed). Colour coding is the same as in the upper panels. 
The inner radius of the truncated accretion disc changes from (a) $100R_{\rm S}$ through (b) $30R_{\rm S}$ to (c) 
$10R_{\rm S}$, replacing the corresponding zones of the hot flow.
Spectra are obtained for the case $\Gamma_{\rm inj}=3.0$. 
}
\label{fig_disc}
\end{figure*}

The turnover frequency of the synchrotron spectrum from each radius is given by equations~(\ref{eq:turn_over_exact}) and 
(\ref{eq:turn_over_sca}).
For the case with $\Gamma_{\rm inj}=2.5$ we substitute parameters of zone~1: $R=10R_{\rm S}$, $B=10^6$~G,
$\tau=10^{-2}$ (Thomson optical depth of the high-energy tail), 
and power-law slopes $\beta=5/4$, $\theta=1/2$ and $p=2$ to obtain the scaling
\begin{equation}\label{eq:turn_over_calc}
 \nu_{\rm t} \approx 10^{15} \left( \frac{R}{10 R_{\rm S}} \right)^{-1} {\rm Hz} . 
\end{equation}
The similar scaling can be obtained for $\Gamma_{\rm inj}=3.0$.
In order to estimate the synchrotron luminosity we substitute the calculated turnover frequency into
equation~(\ref{eq:luminosity})
\begin{equation}\label{eq:luminosity_calc}
 \nu L_{\nu} \approx 2 \times10^{35} \left( \frac{R}{10 R_{\rm S}} \right)^{-7/8} {\rm erg\,s}^{-1},
\end{equation}
which is consistent (within a factor of 2) with the values obtained in precise numerical calculations.

The main model parameters (see Sect.~\ref{sect:setup}) can be constrained by the data. 
The first four parameters ($L$, $\Gamma_{\rm inj}$, $\tau_1$ and $B_1$) can be obtained from the X-ray luminosity and spectral slope, 
the cut-off temperature and the slope of the $\gamma$-ray tail.
The other three parameters ($\theta$, $\beta$ and $R_{\rm tr}$) can then be extracted from the OIR data: 
the turnover frequency (equation~\ref{eq:turn_over_exact}),
spectral slope (equation~\ref{eq:OIR_slope}) and the luminosity (equation~\ref{eq:luminosity}).

The precise values of the minimum and maximum Lorentz factors of the injected power-law electrons do not affect
much the resulting spectra as far as the electrons emitting at the self-absorption frequency remain in a power-law.
We note that very similar results can be obtained by assuming most of the energy goes to heat the thermal distribution
and only a small fraction goes to the power-law tail (see Appendix~\ref{apndx:nth_frac}).
At the same time, the spectra of fully thermal hot flow with the same values of magnetic field are too hard to match the
observations in the hard state.
Assuming $B$ an order of magnitude larger in every zone would produce the spectra reasonably well describing the observed ones. 
OIR spectra in this case can also be described by a power law; however, the turnover frequency in every zone is higher
compared to the case of hybrid electrons (see Appendix~\ref{apndx:nth_frac} for details). Thus, in order to explain the IR
points, a much larger hot-flow size is required.
We also note that purely thermal models are not capable of reproducing the observed non-thermal MeV tails.
 
The considered model qualitatively describes  the spectral properties of the hot flow. 
On the quantitative level, the exact slope of the X-ray spectrum and the relative OIR/X-ray 
luminosities may vary depending on the details of calculations. 
For instance,  the radiative transfer effects (see Appendix~\ref{apndx:transfer}) 
harden a bit the  X-ray spectra of the outer zones, while the OIR spectra are nearly unaffected. 
Also reducing the $H/R$ ratio leads to slightly harder X-ray spectra (see  Appendix~\ref{apndx:transfer}), 
if other parameters ($L$, $\tau$ and $B$) are unchanged. Again, the OIR slope remains the same. 
Thus the model is rather robust in its predicted spectral properties.

\subsection{State transitions}\label{sect:state_trans}

A generally accepted scenario for  the hard to soft state transition involves 
the motion of the cold accretion disc towards the compact object \citep{PKR97,Esin97,Esin98}. 
In this case, the role of the disc increases and it gradually replaces the synchrotron
as a source of seed photons for Comptonization. 
We simulate this action by replacing the spectrum in the corresponding zone of the hot flow 
with a multicolour blackbody disc \citep*{SS73,FKR02} of an appropriate inner radius. 
We take the disc truncation radius $R_{\rm tr}$ equal to the outer radius of the largest zone of the hot flow 
and we keep the outer disc radius at $R_{\rm d,out}=3\times10^4$ $R_{\rm S}$.

The additional seed photons for Comptonization are modelled by the injection of blackbody photons with  temperature 
corresponding to the colour temperature of disc inner radius: 
\begin{equation}
 kT_{\rm col} = 2.3 \left( \frac{L}{L_{\rm Edd}} \right)^{1/4}\!\!
 \left(\frac{3R_{\rm S}}{R_{\rm tr}}\right)^{3/4} 
\!\! \left( 1 - \sqrt{\frac{3R_{\rm S}}{R_{\rm tr}}} \right)^{1/4} \!\!\! \mbox{keV}
\end{equation}
(see Table~\ref{tab_params}).
Given that the transition occurs at almost constant luminosity \citep[see ][]{DGK07},
we assume the luminosity, magnetic field and Thomson optical depth of each hot-flow zone remain the same
as in the hard state (see Table~\ref{tab_params}). 
The relative contribution of the cold disc and the hot flow in the observed spectrum 
depends on the inclination, which we take equal to $60\degr$.
The resulting spectra are shown in Fig.~\ref{fig_disc}.

\begin{figure*}
\epsfig{file=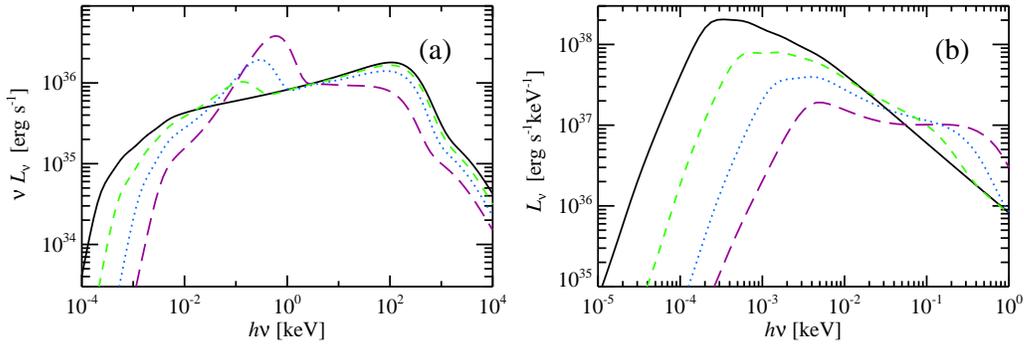, width=14cm}
\caption{
Spectral evolution at state transition: the pure hot-flow spectrum (solid lines), and spectra of the hot flow 
with the cold disc truncated  at $100R_{\rm S}$ (short-dashed), $30R_{\rm S}$ (dotted), and $10R_{\rm S}$ (long-dashed).
(a) spectra in $\nu L_{\nu}$ units and (b) spectra in $L_{\nu}$ units (note the different photon energy range). 
Here the outer disc radius is $R_{\rm d,out}=3\times10^4$ $R_{\rm S}$.
}
\label{fig_transit}
\end{figure*}

\begin{figure*}
\epsfig{file=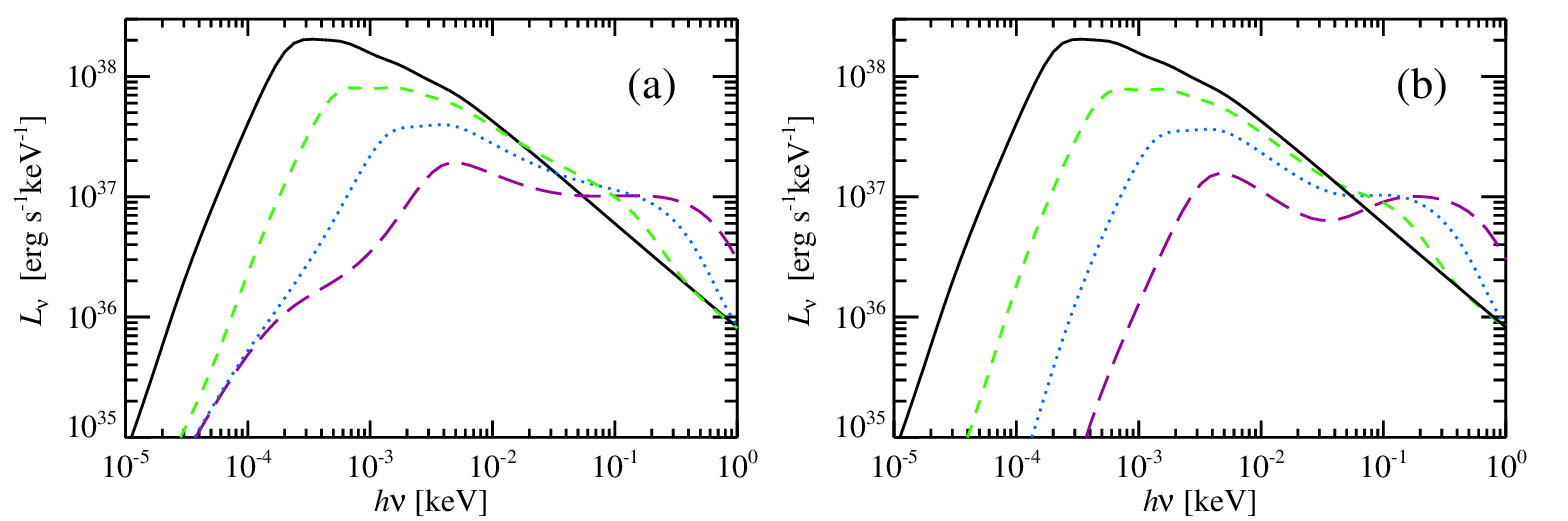, width=14cm}
\caption{
Same as in Fig.~\ref{fig_transit}b, but for outer disc radii (a) $R_{\rm d,out}=10^6R_{\rm S}$ and (b) 
$R_{\rm d,out}=300R_{\rm S}$. 
}
\label{fig_transit2}
\end{figure*}

The total spectrum is now  composed of synchrotron and bremsstrahlung photons, as well as Comptonized synchrotron and disc
radiation. 
The transition between the Comptonization continua from the disc and synchrotron is reflected in the overall spectral
curvature at $\sim$0.1~keV in the spectra of the largest hot-flow zone (e.g., Fig.~\ref{fig_disc}a, short-dashed line).
It is interesting to note that spectra of the zones closer to the BH remain almost unaffected by the disc Comptonization (e.g.,
Fig.~\ref{fig_disc}b, long-dashed line) due to small dilution factor of the disc, while in the outermost hot-flow zone the
cold disc is the dominant source of seed photons.
The cool disc luminosity grows when it moves towards the BH and its photons are much more energetic than those
provided by the synchrotron mechanism; thus, the X-ray spectrum softens as the transition proceeds. 
Once the truncation radius decreases to $10 R_{\rm S}$, the spectrum becomes a combination of the disc and non-thermal
synchrotron, resulting in a softer spectrum with $\alpha\approx -1$.\footnote{Further softening of the spectrum 
is expected if the cold disc penetrates into the hot flow  forming a corona-like geometry \citep{PKR97}. 
This results in additional cooling by the disc photons and reduction of the Compton $y$-parameter.}
As the outburst proceeds, the outer regions of the hot flow collapse
(hours to days before the noticeable X-ray state transition) 
leading to a dramatic drop in luminosity at $\sim$0.1~eV.
The turnover frequency increases and the OIR spectrum becomes harder (see Fig.~\ref{fig_transit}).
Relatively small changes occur around $E\sim30$~eV. 

Hence, one would expect fast change in the luminosity at OIR wavelengths with smaller changes in the UV.
For instance, if one observes the collapse of the $100 R_{\rm S}$ zone while the $30 R_{\rm S}$ zone is still present, there
will be huge changes at $\sim$0.5~eV, while not so significant changes at $\sim$2~eV.
The opposite is expected during the soft-to-hard state transition: when the disc recedes, the hot flow occupies larger and
larger radii and its synchrotron luminosity increases earlier at shorter wavelengths.

Fig.~\ref{fig_transit}(b) shows the $L_{\nu}$ spectra in more details. 
Here we see that the pure hot-flow spectrum below the cut-off in the IR band is a power-law with index $\alpha=5/2$
corresponding to the optically thick non-thermal synchrotron.
We note that such hard spectrum is obtained under the assumption of an absence of the 
hot flow -- disc overlap, i.e. that at distances larger than the hot-flow outer radius the electron density is zero.
In reality, a corona may exist atop of the cold disc, 
thus a gradual transition from the hot flow to the cold accretion disc is
expected leading to the much more gradual turnover of the OIR spectrum.

Fig.~\ref{fig_transit2} illustrates possible spectral features appearing for different sizes of the outer disc radius. 
We see that the hot flow completely dominates the spectrum below $\sim$10~eV if its size  is larger than $100 R_{\rm S}$. 
In this case, the exact value of $R_{\rm d,out}$ does not play any role (unless reprocessing in the outer disc starts to be
important, see Sect.~\ref{sect:irrdisc}).
The largest changes occur for smaller truncation radius and large $R_{\rm d,out}=10^6 R_{\rm S}$. 
For such large discs (see Fig.~\ref{fig_transit2}a), radiation in the far IR is dominated by the Rayleigh-Jeans part of the
spectrum from the outer cold disc. 
The UV radiation is mostly produced at the inner disc edge. 
Synchrotron from the hot flow is still important in the optical.

\section{Comparison with observations}\label{sect:compare_jet}
\label{sect:comparison}

In the present work we considered an inhomogeneous hot accretion flow model for the broadband spectra of the 
accreting BHs. In the hard state, when the standard cold  disc is truncated at a large radius, the 
central hot region is radiating mostly via thermal Comptonization of the non-thermal  synchrotron photons.
Hot flow extending over a large range of radii produces a power-law-like flat spectrum  in the OIR range.

In the soft state, the cold disc moves in, brightens and takes over as a source of seed soft photons for Comptonization. 
This effectively reduces the role of synchrotron radiation in electron cooling. 
At the same time, the reduction of the size of the emitting region results in an increase of the synchrotron 
self-absorption frequency, making the synchrotron emission in the IR band negligible.  

The scenario considered in this work is capable of reproducing the broadband spectra  from the 
IR to the gamma-rays of BHs in all spectral states. 
However, the spectral data  alone are not capable of distinguishing among various models and they have to be considered
together with other sources of information (e.g. timing and polarization).
Below we will discuss in details various properties of the developed hot-flow model and compare them to observations. 
We also compare our model to the popular jet scenario.

\subsection{Hard state}

\subsubsection{X-ray spectrum and variability}\label{sect:obs_hard_state_X}

In our model the X-ray spectrum is dominated by the Comptonization continuum from the innermost zone 
where most of the gravitational energy is dissipated. 
In the X-ray range it  can generally be described by a power law.
Outer zones of the hot flow have softer spectra, because of a larger role of 
non-thermal synchrotron and the increasing amount of the cold disc photons. 
The overall spectrum is thus slightly concave. 
Such spectra are consistent with those observed from the BHBs. 
For example, the best studied BH, Cyg~X-1, clearly has a concave spectrum \citep{FPZ01}
that can be fitted with two Comptonization continua \citep{IPG05}.

A larger contribution from the outer zones to the soft X-rays should be reflected in the variability properties. 
Assuming that variability is produced by propagation of fluctuations in the mass accretion rate 
through the disc \citep{Lyub97,KCG01},  we expect an increase in  the variability amplitude for higher 
photon energies at higher Fourier frequencies, which  is indeed observed \citep{Nowak99a}. 
The autocorrelation function of soft X-rays in our model is expected to be wider than that of the hard X-rays, 
consistent with what is measured in Cyg~X-1 \citep{MCP00}.
The same effect is more obviously seen in the Fourier-frequency-resolved spectra \citep*{RGC99,GCR00}, 
which are softer and have larger reflection amplitude at low Fourier frequencies. 
This implies that the outer zones of the hot flow (which are closer to the cold reflecting medium)
give a relatively larger contribution to the soft X-ray flux than to the hard X-rays. 
The reduction of the equivalent width of the 6.4~keV Fe line in the frequency-resolved spectra above 1~Hz
suggests that the cold disc truncation radius is of the order of 100 $R_{\rm S}$
\citep{RGC99,GCR00}, further supporting our scenario. 
Similarly, even larger inner radii of the cold disc were measured in the low-extinction BH transient XTE~J1118+480
\citep{Esin01,CHM03}.

Another important finding is that the harder X-rays are delayed with respect to the soft X-rays 
(\citealt{Nowak99a}; \citealt*{Nowak99b}). 
The large values of these hard time lags and their frequency-dependence $f^{-1}$ 
can naturally be explained by spectral pivoting of a power-law-like spectrum \citep{PF99,Pou01}. 
The spectral evolution can arise when the accretion rate fluctuations 
propagate towards the BH into the zone with harder spectra  \citep{KCG01}, 
again consistent with our multizone hot-flow model.

\begin{figure*}
\epsfig{file=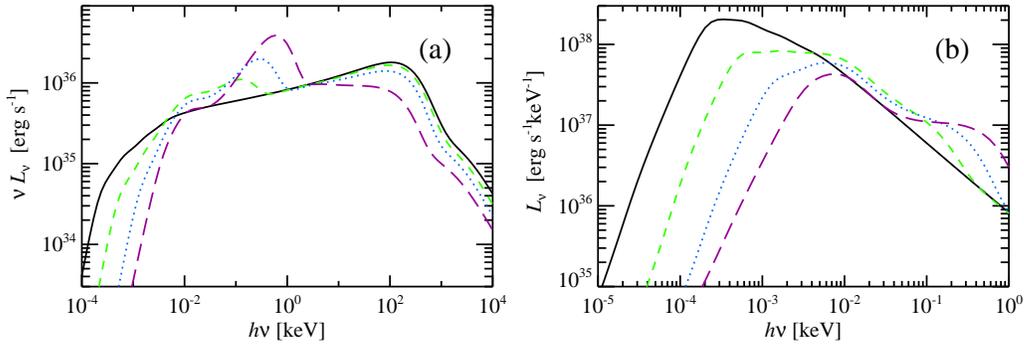, width=14cm}
\caption{
Same as in Fig.~\ref{fig_transit}, but with the additional contribution from the irradiated accretion disc.
}
\label{fig_irrad}
\end{figure*}

\subsubsection{OIR excesses and flat spectra}

The OIR excesses above the standard disc spectrum were reported in a number of sources: XTE~J1859+226 \citep{HHC02}, 
XTE~J1118+480 \citep{HHC00,Esin01,CHM03}, GX~339--4 \citep[e.g.,][]{GBRC11,SUT11}, A0620--00 \citep{GMM07}, 
SWIFT~J1753.5--0127 \citep{CDSG10}, V404 Cyg \citep{HBR09}.
In our model, the OIR spectrum consists of two components (Fig.~\ref{fig_disc}): one comes from the multicolour accretion disc
and another from the hot flow.
The relative role of these components varies with the wavelength (Figs~\ref{fig_transit}~and~\ref{fig_transit2}).
The disc spectrum is hard in the OIR band, while the non-thermal synchrotron from the hot flow is typically softer with
$\alpha_{\rm OIR}\sim 0$.
The second component thus produces an excess emission.

In many cases the contribution of the non-thermal component is rather small compared to the disc, and 
it can be seen only as the IR excess.
On the other hand, sometimes the synchrotron component dominates, which results in an almost pure power-law OIR spectrum.
A good example is XTE~J1118+480, where the spectral index $\alpha_{\rm OIR}=-0.15$ was measured \citep{CHM03}.
This spectral index can be reproduced in our model, for example, with parameters $\theta=0.5$, $p=2.0$ and $\beta=5/4$ 
(Fig.~\ref{fig_pureSSC_inj2_5} illustrates this case).

\subsubsection{Optical/X-ray cross-correlation}

In recent years, a number of simultaneous optical(IR,UV)/X-ray observations with high time resolution were performed
\citep{Kanbach01,HHC03,GMD08,HBM09,DSG11}, all revealing the intrinsic connection of the two light curves on subsecond
time-scales.
The computed CCF have a complicated shape with a dip in the optical light curve preceding the X-ray
peak (the so-called precognition dip), together with an optical peak lagging the X-rays.

The behaviour can be explained if the optical emission consists of two components: one coming from the synchrotron in
the hot flow and another from reprocessed X-ray emission \citep{VPV11}.
Increase of the mass accretion rate causes an increase in the X-ray luminosity and affects the parameters of the hot
flow, leading to a higher synchrotron self-absorption.
The latter results in a drop of the optical emission, therefore these two energy bands appear anticorrelated.
This is reflected in the negative CCF with the shape resembling that of the X-ray autocorrelation function. 
On the other hand, the reprocessed radiation is delayed and smeared, giving rise to a CCF peaking 
at positive lags (optical delay). The combined CCF has a complicated shape consistent with the data. 
From the point of view of the multizone consideration, with a small increase of the mass accretion rate the cool disc moves inwards 
and causes the collapse of the hot flow at large radii.
Thus, the suppression of the OIR emission with increasing X-ray radiation is also expected in this scenario.

A different, one-peak structure of the IR/X-ray CCF  was found in GX~339--4 \citep{CMO10},  
suggesting the hot flow was not the dominant source of the correlated variability during their observations.
However, it might still give significant contribution to the constant flux component, but less to the varying component,
and therefore would not be detected in the timing analysis.
As it can be seen from Figs~\ref{fig_pureSSC_inj2_5}~and~\ref{fig_pureSSC_inj3}, 
the zones giving major contribution to the IR wavelengths do not contribute much to the X-rays. 
Therefore, the fraction of the correlated variability coming from these regions is expected to be small, and  
another source (likely the jet emission, as suggested in \citealt{CMO10}) might be responsible for the shape of
the CCF.
We also note that large amplitude of fluctuations in the IR light-curve suggests that the source of correlated
variability is also dominating over the constant flux component.

The optical correlation with the X-rays was also detected in the quiescent state of V404 Cyg \citep{HBR09}, while no clear
radio/X-ray (nor radio/optical) correlation was found on the time-scales of hours, again suggesting the radio and optical/X-ray
emission are  produced by the different components.

\subsubsection{Irradiation of the cold disc}
\label{sect:irrdisc}

The X-ray radiation from the hot flow can be intercepted and reprocessed in the cold disc. 
The irradiation strongly depends on the disc outer radius and the disc shape.
The larger is $R_{\rm d,out}$, the cooler can be this emission.
The more flared is the disc, the larger is the reprocessed luminosity, which typically is expected to give significant
contribution to the OIR band, exceeding the viscous disc luminosity at these energies \citep{SS73}.
Presence of the irradiated disc can be also reflected in the X-ray time-lags \citep[see][and reference therein]{Pou02} and 
in the optical/X-ray CCF \citep{VPV11}. 
Its signatures are also seen in the spectrum \citep[e.g.,][]{HHC02,GDP09}.

For typical parameters of LMXBs with the disc size of $10^{11}$~cm, the X-ray luminosity of $10^{37}$~erg~s$^{-1}$ and
10 per cent reprocessing efficiency, the temperature of the outer disc is about 1.2~eV.
For an illustration, we have added the emission from the irradiated discs to our hot-flow spectra (see Fig.~\ref{fig_irrad}).
Following \citet{Cun76}, we assumed the following dependence of the effective temperature on radius
\begin{equation}
 kT_{\rm irr} = 1.2 \left( \frac{R}{R_{\rm d, out}}\right)^{-3/7} \mbox{eV} 
\end{equation}
and the outer disc radius of $R_{\rm d, out}=10^{11}$~cm.

As can be seen in Fig.~\ref{fig_irrad}, for typical parameters of the BHBs the spectrum of the irradiated disc peaks
around 5~eV and has a Rayleigh-Jeans-like tail with $\alpha\sim2$ in the OIR.
If the inner hot-flow size exceeds 30--100~$R_{\rm S}$, its synchrotron emission will dominate over the
reprocessing below $\sim$1~eV.
We do not expect much of reprocessed emission below 0.5~eV, except for the large-period systems (similar to V404~Cyg and
GRS~1915+105).
Therefore, at sufficiently long wavelengths in scenario of \citet{VPV11} we expect the IR/X-ray cross-correlation
function to have only the dip, with no peak.
The overall spectral shape is complex, with a number of bumps corresponding to the standard disc (above $\sim$100~eV), the
irradiated disc and synchrotron from the hot flow (in the 1--10~eV range).
A hardening of the spectrum observed in GX~339--4 at $\sim$2--3~eV \citep{Buxton12,DKB12,RCC12} can be due to a
transition from the hot flow to the irradiated disc spectrum.

\subsection{State transition}\label{sect:comparison_trans}

\subsubsection{Broadband spectral evolution at state transition}

During the hard-to-soft state transition the outer zones of the hot flow gradually collapse.
This leads to a drop in luminosity at longer wavelengths preceding the subsequent drop at shorter wavelengths.
The corresponding time delay between sharp luminosity changes at different wavelengths corresponds to the viscous
time-scale of the cold disc between the corresponding radii and, depending on the separation of the wavelengths and accretion
parameters, can be as short as hours (e.g., if one observes in different optical filters) and as long as a few days (e.g., IR
and UV).
The dramatic changes in $V$, $R$ and $I$ bands were detected, e.g. in GX~339--4 during the rising phase of the 2010 outburst
\citep{CB11}; however, the time resolution of the light-curves does not allow to judge on the exact time of flux quenching at
different wavelengths.
A much more gradual decrease of the UV flux before the X-ray spectral transition \citep{YY12} is
in principle consistent with our model if one also considers contribution from the irradiated disc (e.g., as in
Fig.~\ref{fig_irrad}: gradual decrease at $\sim$10~eV, with sharp drop at $\sim$1~eV).
During the reverse (soft-to-hard) transition, the fast luminosity increase at shorter wavelengths is expected to occur earlier
than at longer wavelengths, and the corresponding time delays are expected to be larger than in the rising phase, 
as was seen  in the 2005 outburst decay of GX~339--4 \citep{CCB09}.

Recently, the entire transition of the BH transient XTE~J1550--564 from hard to the soft state and back was monitored 
in the $V$, $I$ and $H$ filters \citep{RMDF11}.  
Just before the X-ray spectral transition as well as after the reverse transition, 
significant colour variations occurred, as indicated by the rapid changes in the $H$ band
at  almost constant $V$ magnitude.
In terms of our model, the observed colour change is related to the collapse/recover of a 
zone in the hot flow that is responsible for the $H$-band emission (see Section~\ref{sect:state_trans}).
As is known, the hard-soft and the soft-hard spectral transitions occur at different X-ray luminosities 
\citep[e.g.][]{ZGM04}. This hysteresis most probably is related to the fact that  at the same luminosity 
the cold disc is further away from the central source on the rising phase of the outburst than on the decline. 
According to our model, the hysteresis has to be reflected also in the OIR spectra, namely 
the fast colour change  should occur at a higher X-ray luminosity on the  rising phase than on the decline. 
This is indeed observed  \citep{RMDF11}. 

In the soft state, the accretion disc extends to the last stable orbit, leaving no possibility for the inner hot flow to exist. 
However, the corona should still  be present as supported by the existence of the non-thermal tails 
(produced by the inverse Compton scattering of the disc photons).  
It may also produce synchrotron radiation in the OIR band, but at a much lower level.

\subsubsection{Change of the X-ray radiation mechanism}

At luminosities above a few per cent of Eddington, BHBs show a strong correlation between spectral index and luminosity.
At lower luminosities the trend is reversed  \citep{SPDM11}. 
Similarly, an indication of the reverse trend was detected in low-luminosity AGNs \citep{Const09}. 
This was interpreted as a change of the source of seed photons for Comptonization from the disc photons 
dominating at higher luminosities to the synchrotron at lower luminosities.  
The anti-correlation at $L/L_{\rm Edd}\sim 10^{-3}$--$10^{-2}$  
can  be reproduced within a two-temperature hot accretion flow model \citep*{NXZ12}. 
The whole spectral   index -- luminosity dependence is well explained by   
one-zone hybrid Comptonization model \citep[see figs~7 and 12 in ][]{VVP11}. 
The multizone consideration presented in the current paper follows the same pattern as 
the one-zone model, because the X-ray spectrum is dominated by radiation from the inner zone.

\subsection{Polarization}

The only indication of the X-ray polarization from BHB goes back to the OSO-8 satellite 
\citep{Weisskopf77}, which measured 3.1$\pm$1.7 per cent linear polarization from Cyg X-1 at 2.6~keV.  
Such a polarization can be produced by Compton scattering if the geometry of the X-ray emitting region is a
flattened disc-like structure \citep[$H/R\sim0.2$ according to the calculations of][]{LS76}.
The number of scatterings the X-ray photons undergo 
depends on the electron and the seed photon temperatures. 
For a 100~keV plasma, photons double their energy in each scattering: hence, the disc photons 
of a typical energy of 0.5~keV would reach 3~keV in only three scatterings, 
while the synchrotron photons emitted at 10~eV require about eight scatterings. 
Thus even if the synchrotron photons are polarized, this information 
is forgotten, and the X-ray polarization is completely determined by the geometry of the medium.
In our model, we considered the case of $H/R=0.5$, however, the spectral shape remains the same even for the flatter 
geometry (see Appendix~\ref{apndx:transfer}). Thus, the polarization measurements are consistent with the hot-flow model.

Recently, strong linear polarization ($\Pi=67\pm30$ per cent) 
in the soft  $\gamma$-rays above 400 keV was detected in Cyg X-1 with the IBIS instrument onboard {\it INTEGRAL} \citep{LRW11}.  
Similar polarization  ($\Pi=76\pm15$ per cent)  was also observed with the SPI spectrometer \citep{JRCC12}. 
The polarization angle of $40\degr-42\degr$ is about $60\degr$ away 
from the radio jet axis at $\approx-20\degr$ \citep{JRCC12,ZLS12}.  
Such a large polarization degree in the MeV range is extremely difficult to get in any scenario. 
Synchrotron jet emission from non-thermal electrons in a highly ordered magnetic field 
can have a large polarization degree (up to $\sim$70 per cent) in the optically thin part of the spectrum, and indeed 
a high polarization in the radio and the optical bands reaching 30--50 per cent is observed from extragalactic relativistic jets 
(\citealt*{Impey91}; \citealt{Wills92,Lister01,Marscher02,Ikejiri11}). 
However, this scenario also needs a very hard electron spectrum as well as an extreme fine-tuning to 
reproduce the spectral cutoff at a few MeV  \citep{ZLS12}.  
In the hot-flow scenario, the MeV photons are produced by non-thermal Compton scattering of the  100 keV photons by 
electrons with $\gamma$$\sim$2--4. 
These electrons cannot be isotropic, because no significant polarization is expected  in that case \citep{Pou94ApJS}.  
This then implies that they must have nearly one-dimensional motion, e.g. along the  large-scale 
magnetic field lines threading the flow. 
The $60\degr$ offset of the polarization vector relative to the jet axis then implies the inclined field lines. 
If the measured high polarization degree is indeed real, 
this would put strong constraints on the physics of particle acceleration in the hot flow 
and the magnetic field geometry. 
 
The polarization degree of the hot-flow radiation in the OIR band is strongly affected by the Faraday rotation. 
The rotation angle $\chi_{\rm F}\sim 10^6 \tau B_{||,6}  \nu_{15}^{-2}$ exceeds $10^5$ rad 
and the polarization degree is expected to be essentially zero in the optically thin part of the spectrum. 
In the optically thick regime, the intrinsic flow polarization (parallel to the field lines) 
is not more than about 10 per cent even for  the ordered magnetic field and 
without Faraday rotation \citep{PS67,GS69}. Thus the OIR polarization is expected to be very low. 

Optical and UV radiation from the cool accretion disc may also be polarized up to $\sim$11.7 per cent (parallel to the disc plane)
at large inclinations, if the opacity is dominated by the electron scattering \citep{Cha60,Sob63}. 
At lower energies, absorption in the atmospheric layers of the disc \citep{LS81,LS82} 
and the  Faraday rotation reduce the polarization degree and it can drop down to zero in the IR. 

A detection of linear polarization at a few per cent level in the OIR bands in BHBs 
(\citealt{SHH04,SFW08,RF08}; \citealt*{CDR11}) is consistent with 
being produced either by the jet synchrotron radiation, extended photosphere of the hot flow, 
dust/electron scattering in the source vicinity or by the interstellar dust.

\subsection{Comparison with the jet paradigm}

The simplest jet model has a conical geometry with all parameters distributed as a power law with distance and 
with electrons having  a power-law distribution in Lorentz factor \citep{Marscher77,BK79}.
The magnetic field can be assumed ordered or tangled, but this only influences the polarization properties. 
The mathematical formulation of this model is identical to the analytical model 
for the hot flow considered in Sect.~\ref{sect:model}.
Such simplified jet model was applied to the broadband spectra of BHBs \citep*{MFF01}.
In a more realistic situation, the electron distribution would be subject to acceleration mechanisms, as well as cooling
processes such as Compton, synchrotron and adiabatic \citep[as e.g. in][]{PC09,PM12}. 
However, due to a complicated physics, the energy input and the acceleration efficiency throughout the jet are generally
unknown, and in the spectral models remain ad hoc functions.
Thus  the shape of the resulting spectra and its total energetics strongly depend on the assumptions.
This disadvantage is avoided in the hot-flow models: the input energy here comes from the liberated gravitational energy, 
which can be estimated analytically, and the acceleration efficiency (as well as its role compared to other heating mechanisms)
does not significantly affect the final spectrum.
Both scenarios, however, suffer from a large number of parameters due to an absence of first principle model. 

Apart from the different physical assumptions, the jet and the hot-flow models predict different spectral properties.
In the simple jet model, the synchrotron spectrum consists of the optically thin part with spectral slope 
$\displaystyle \alpha=-(p-1)/2$ and the optically thick part (sum of contribution from different zones) 
with the same slope as given by equation~(\ref{eq:OIR_slope}).
The low-energy cut-off is determined by the jet extension and falls in the radio wavelengths. 
In contrast, the cut-off of the hot-flow spectrum is related  to the truncation  radius of the disc 
and likely falls in the OIR band.
The break energy to the optically thin part of the jet is determined by the extension of the injection zone and  
for X-ray binaries is expected to fall in the IR wavelengths \citep{HS03},
while in the hot-flow model the break is in the UV/optical band as determined by the size of the inner region.
The jet optically thin synchrotron is sometimes claimed to extend to the X-rays, while in the hot-flow model 
radiation at these energies is produced in the Comptonization processes.

Let us  now assess possible contribution of the jet and the hot flow to various wavelengths relying on the observed
spectral properties.
The X-ray spectra of BHBs have sharp cutoffs at about 100~keV that are impossible to produce by 
non-thermal synchrotron even if the electron distribution has an abrupt cutoff \citep{ZLG03}. 
The hard spectra are also difficult to produce by optically thin synchrotron as this requires 
a very hard injection. 
Another question is then about the observed low level of the X-ray polarization, which is in contradiction with the theoretical
expectations from the optically thin synchrotron, as well as with the levels measured in extragalactic sources.
On the other hand, the X-ray spectral properties are well explained by the (nearly) thermal Comptonization in the hot
flow (see Sect.~\ref{sect:obs_hard_state_X}).

Atop of the X-ray power-law, the Compton reflection and the iron line are rather often detected features.
Their amplitude and its correlation with the underlying spectral shape
strongly argue in favour of the small X-ray emission region and against any beamed-away radiation from the jet. 
The analysis of the hard state X-ray spectra of the BHB Cyg~X-1 revealed that 
the spectra pivot at energies of 10--50~keV \citep{ZPP02}.
If the entire IR-to-X-ray continuum is produced by the same region (as proposed in the jet model), 
such pivoting would predict two orders of magnitude variations at 1~eV in the hard state, which are not observed. 
On the contrary, X-ray pivoting can be easily understood in the hot-flow model, where it can be produced by 
small variations of the cold disc radius and varying injection rate of the disc photons (see Fig.~\ref{fig_irrad}). 

The observed fast X-ray variability and hard time lags can naturally be understood in the hot-flow model where
the X-ray emitting region is small, and the lags are related to the viscous time-scale of propagating fluctuations
\citep{KCG01}.
At the same time, the lags can occur from Compton scattering delays within the jet \citep{Kylafis08}, however, 
this model contradicts the observed narrowing of the auto-correlation function with energy \citep{MCP00}.

The MeV tails detected in spectra of a number of hard-state BHBs are likely produced by the non-thermal electrons, either by
the optically thin synchrotron emission (as in the jet) or by the inverse Compton scattering (as in the hot flow).
The detailed investigation of this tail showed that it can be explained by the high-energy end of the jet synchrotron
emission, however in this case one needs to assume a very hard index of the electron power-law distribution $p=1.3-1.6$,
which is in conflict with standard acceleration models and observations \citep*[see][and references therein]{ZLS12}.
At the same time, the hot flow easily accounts for the MeV tails in the hard state as well as in the soft state 
(where it can be replaced by a corona), when the jet is quenched.

The non-thermal OIR radiation has often been interpreted as a jet emission \citep[see review by][]{RF09}.
The similarities in the IR and radio light-curves in microquasars, such as GRS~1915+105 \citep{FPB97}, 
indeed favour common source of variability.
However, this scenario meets substantial problems in a number of other sources.
For instance, in some systems the jet broken power-law model, normalized to fit the radio fluxes, significantly underpredicts
the optical luminosity even after accounting for possible irradiated disc contribution \citep{Soleri10,CB11}. 
Sometimes the OIR slope is different from the radio as, for example, in XTE~J1118+480 the OIR spectrum 
with $\alpha_{\rm OIR}=-0.15$ is much softer than the radio spectrum with $\alpha_{\rm R}=0.5$  
\citep{HMH00,CHM03}, but much harder than expected from the optically thin jet emission. 
The hot-flow scenario can reproduce the observed flat OIR spectra, at the same time the slopes in the OIR and radio do not
necessarily match, as they are produced in different regions (inflow and outflow).

A rather low OIR polarization (at most a few per cent) observed in BHBs 
is clearly much below the high polarization observed in many extragalactic jets.
On the other hand, certain types of objects, the so-called compact steep-spectrum and gigahertz peaked-spectrum radio sources,
demonstrate similarly small polarization levels (0.2 and up to 7 per cent, respectively, \citealt{ODea98}).
However, the (rest-frame) break frequency measured in these objects is quite low; thus,
in the  X-ray binary jets, as expected from the scaling laws the break frequency  is in the (sub-) millimeter range.

The two models make different predictions for changes of the OIR spectrum during the state transitions.
In the hot-flow scenario, the spectrum gradually hardens at hard-to-soft transition on the time-scales of days to weeks,
corresponding to the typical time-scales on which the cold accretion disc evolves.
It softens again at the reverse transition.
On the other hand, one would not expect systematic changes of the jet spectral slope, as any fluctuations
would propagate through the jet on time-scales of hours, much shorter than the state transition.
However, if the jet power is gradually decreasing at the transition, then the dependence of the turnover frequency on the
mass accretion rate suggests that $\nu_{\rm t}$ is also decreasing \citep{HS03}.
Thus, the jet scenario predicts softening of the OIR non-thermal spectrum at hard to soft state transition and hardening at the
reverse transition, opposite to the hot-flow scenario.

In reality, the OIR emission can contain contributions from a number of components: the hot flow, the jet, 
and the cold accretion disc (likely irradiated).
It is also possible that in some sources there is a dip in the microwave band, where the transition from the radio jet
to the hot flow occurs.
The dip can be detected in the far infrared/submillimeter wavelengths, 
which were not systematically studied in the past.
It is thus of high interest to observe at these wavelengths, especially with the available capabilities of 
the Atacama Large Millimeter/submillimeter Array.

\section{Summary}
\label{sect:summary}

The observed OIR flat spectra and the MeV tails evidence the significant role of non-thermal electrons in spectral
formation of accreting BHBs. 
On the other hand, the commonly detected X-ray spectral cut-offs at $\sim$100~keV can be produced only by thermal particles.
Whether these two populations belong to one component or originate from completely different places is debated.
We present a model, in which the entire infrared to X-ray/$\gamma$-ray continuum is produced by one component,
the inhomogeneous hot accretion flow, present in the vicinity of compact object.
The difference from the earlier studied hot geometrically thick optically thin flows is that 
the steady-state electron distribution in our model is hybrid, i.e. Maxwellian with a weak high-energy tail.

The X-ray spectra in our model are dominated by the radiation of the innermost regions of the hot flow.
For this reason, the model inherits the advantages of the one-zone synchrotron self-Compton model
\citep{MB09,PV09} that explains well the X-ray spectral properties of BHB in their hard state, such as
\begin{enumerate}
\item stable spectra with photon index $\Gamma\sim$1.6--1.9 and the cutoff at $\sim$100~keV in the hard state,
\item low level of the X-ray polarization,
\item presence of the MeV tail in the hard state,
\item power-law-like X-ray spectra extending to a few MeV in the soft state,
\item softening of the X-ray spectrum with decreasing luminosity below $\sim$10$^{-2}L_{\rm Edd}$,
\item weakness of the cold accretion disc component in the hard state, and 
\item correlation between the spectral index, the reflection amplitude, the width of the iron line and 
      the frequency of the QPO.
\end{enumerate}
We show that the multizone consideration allows to understand many other observables in the context of the hot-flow model:
\begin{enumerate}
\item hard X-ray lags with logarithmic energy dependence,
\item concave X-ray spectrum,
\item non-thermal OIR excesses and flat spectra,
\item strong correlation between OIR and X-ray emission and a complicated shape of the CCF, and  
\item a complex evolution of the OIR--UV spectrum during the state transition.
\end{enumerate}

We present relevant analytical expressions to estimate the hot-flow parameters from the OIR data
(see eqs~\ref{eq:turn_over_exact}, \ref{eq:OIR_slope} and \ref{eq:luminosity}). 
Additional X-ray and $\gamma$-ray data are required to find a complete parameter set.
However, the hot flow extent can be found from the OIR data alone under certain assumptions
(eqs~\ref{eq:turn_over_calc} and \ref{eq:luminosity_calc}).

We compare the developed model to the popular jet scenario and show that in a number of cases the data favour 
the hot-flow interpretation.
We encourage future observations in the far infrared and submillimeter wavelengths to provide  the 
missing link between radio and infrared, which would allow us to determine the contribution 
of the two components to the OIR emission. 

\section*{Acknowledgments}

The work was supported by the Finnish Graduate School in Astronomy and Space Physics (AV), 
the Academy of Finland grant 127512 (JP) and ERC Advanced Research Grant 227634 (IV).
We thank Andrzej Zdziarski for fruitful discussions and useful comments, 
Marion Cadolle Bel and Dave Russell for conversations, which helped us to make the paper more understandable.
We also thank anonymous referee for many useful suggestions, which helped us to better justify our model and to improve the
paper.


\appendix

\section{Typical time-scales}\label{apndx:times}

\subsection{Radiative versus Coulomb time-scales}\label{apndx:rad_coul_times}

The synchrotron cooling time can be calculated as \citep[e.g.][]{RL79}
\begin{equation}\label{eq:time_s}
 t_{\rm cool, s} = \frac {\gamma - 1}{|\dot{\gamma}_{\rm s}|}
                  =     \frac {1}{\gamma + 1} \left( \frac{4}{3} \frac {\sigma_{\rm T} U_{\rm B}}{m_{\rm e}c} \right)^{-1},
\end{equation}
where $\dot{\gamma_{\rm s}}$ is synchrotron cooling rate.
The cooling time for Compton scattering is similar.
After a little algebra one can obtain
\begin{equation}
 t_{\rm cool, s}  \approx \frac{7.5\times10^{-4}}{\gamma+1} \frac{R}{c \eta_{\rm B}} 
                        \left( \frac{L}{L_{\rm Edd}}\right)^{-1} \left( \frac{R}{10 R_{\rm S}}\right),
\end{equation}
where $\eta_{\rm B}=3U_{\rm B}/(4\pi U_{\rm rad})$ denotes the ratio of the magnetic and radiation field energy densities.
The typical time-scale of electron-electron Coulomb energy exchange at the equilibrium can be estimated as \citep[e.g.][]{NM98}
\begin{equation}
 t_{\rm cool, Coul} = \frac {\gamma - 1}{|\dot{\gamma}_{\rm Coul}|} \approx
                    \frac {2 R}{3 c} \frac{\overline{\gamma}_{\rm eq}} {\tau \ln \Lambda}
                    \frac {z^3}{\gamma(\gamma+1)},
\end{equation}
where $z$ is the particle momentum in units of $m_{\rm e}c$, 
$\overline{\gamma}_{\rm eq}$ is the average Lorentz factor of the electrons in equilibrium
and $\ln \Lambda$ is the Coulomb logarithm.
At higher Lorentz factors the cooling is determined by radiative processes, while for lower $\gamma$ the non-radiative Coulomb
collisions are dominant.
The relative role of radiative and Coulomb cooling changes with radius:
\begin{equation}
 \frac{t_{\rm cool, Coul}}{t_{\rm cool, s}} \approx 56 \frac {z^3 \eta_{\rm B}}{\gamma \tau}
                                                 \left( \frac{L}{L_{\rm Edd}}\right)\left( \frac{R}{10 R_{\rm S}}\right)^{-1},
\end{equation}
where we used $\overline{\gamma}_{\rm eq}=1$ and $\ln \Lambda = 17$.
Therefore, the Coulomb exchange rates start dominating over the radiative cooling with an increasing size.

\subsection{Coulomb versus accretion time-scales}\label{apndx:adv_times}

We assume the photon spectra and electron distributions in each zone are in equilibrium; thus, the typical time-scales of
equilibration should be much less than the dynamical time-scale at a given distance.
To estimate the latter, we consider the properties of an advective hot flow, derived by \citet{NY94}.
Taking their radial velocity approximation, viscosity parameter 0.1 and assuming the adiabatic index 3/2 \citep{QN99}, we
obtain the advection/accretion time
\begin{equation}
 t_{\rm adv} \approx 20 \sqrt{\frac{R^3}{GM}}.
\end{equation}
We can compare the Coulomb (the radiative cooling is faster than Coulomb for $R\le100R_{\rm S}$) and advection time-scales
(again, assuming $\overline{\gamma}_{\rm eq}=1$)
\begin{equation}
 \frac{t_{\rm cool, Coul}}{t_{\rm adv}} \approx 5\times10^{-4}\frac{\gamma}{\tau} 
                                                \left( \frac{R}{10 R_{\rm S}} \right)^{-1/2}.
\end{equation}
Hence, for typical parameters $\gamma\sim$1--100 and $\tau\sim$0.1--1 the cooling time-scale is  
shorter than the advection time.

\section{Modifications of the original model} 
\subsection{Non-thermal fraction}
\label{apndx:nth_frac}

One of the main assumptions in our work is that all the energy dissipated in the flow is transferred to the electrons by
acceleration, i.e. 100 per cent of the power is injected in the form of power-law electrons.
In reality, many other dissipation mechanisms may play a significant role in particle heating, such as electron-proton Coulomb
collisions (the main energy transfer mechanism in the advective flow models), 
resonant interactions with plasma waves and other collective plasma effects.
These acceleration processes do not lead to power-law electron distributions; 
instead, the Maxwellian distribution is heated as a whole, achieving higher temperature.
Thus it is important to check how robust the model is  to changes in the fraction of non-thermal particles.  

\begin{figure}
\epsfig{file=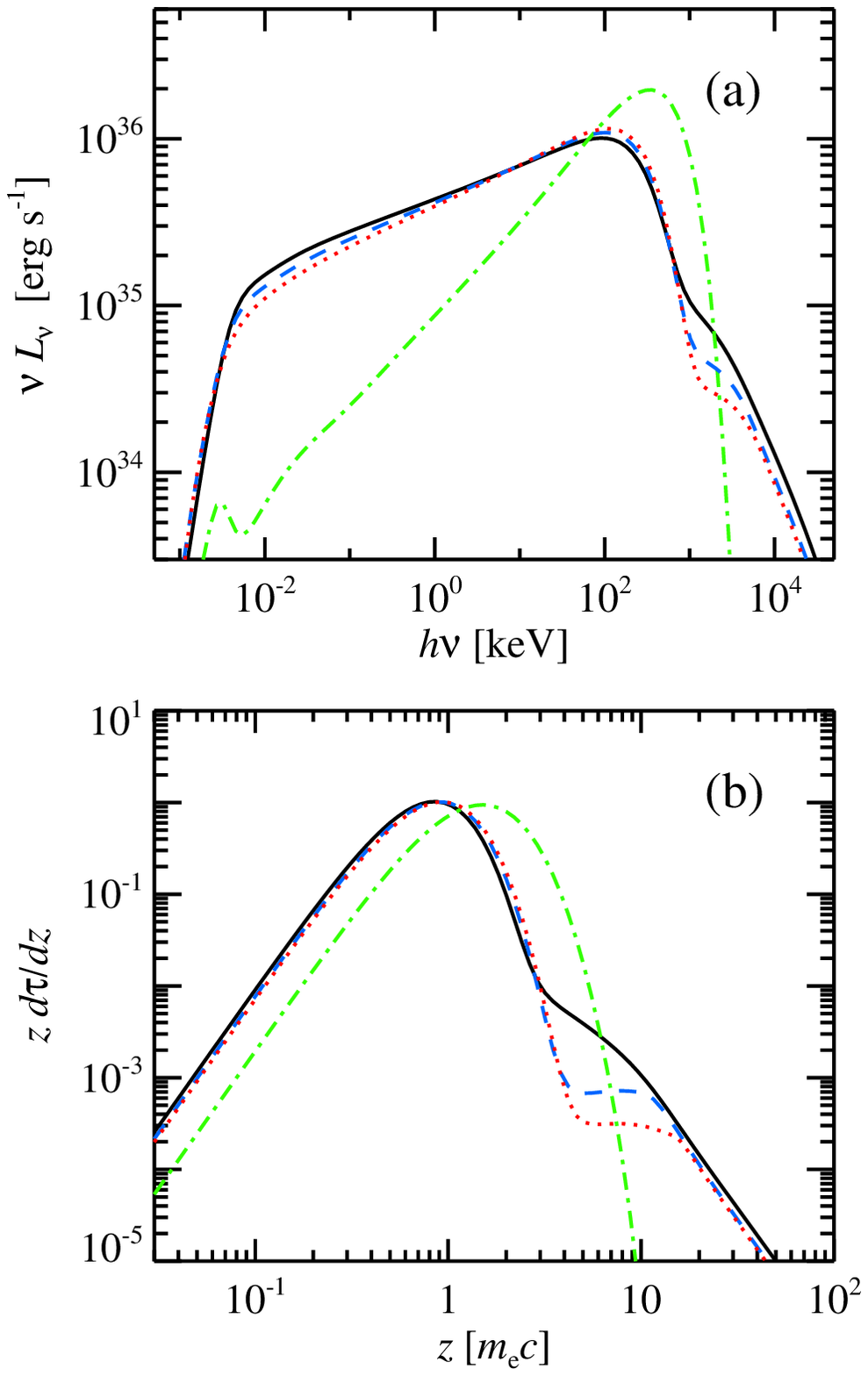, width=7cm}
\caption{
Effects of decreasing fraction of energy given to acceleration.
Panel (a) shows resulting spectra and panel (b) gives the electron distributions.
Solid black line corresponds to $f_{\rm nth}=1.0$ (same as purple long-dashed line in Fig.~\ref{fig_pureSSC_inj3}), 
$f_{\rm nth}=0.19$ (blue dashed), $f_{\rm nth}=0.13$ (red dotted) and $f_{\rm nth}=0$ (thermal particles, green
dot-dashed). For power-law electrons the injection index $\Gamma_{\rm inj}=3.0$ was assumed. 
}
\label{fig_nth_frac}
\end{figure}

We note that effects of decreasing non-thermal injection fraction were considered in \citet[][fig.~9]{MB09}, who showed that as
long as this fraction is more than $\sim$30 per cent, the results are very similar.
However, they fixed the minimum and the maximum Lorentz factors of the injection; 
thus, decreasing the non-thermal fraction would
result in decreasing of normalization of the injected power-law electrons.
Here we investigate the effects of decreasing non-thermal energy fraction 
keeping the normalization of the injected power-law constant by increasing $\gamma_{\rm min}$.
As previously mentioned, the maximum Lorentz factor of accelerated electrons $\gamma_{\rm max}=10^3$ is assumed.
The energy fraction given to power-law electrons is calculated as
\begin{equation}
 f_{\rm nth} = \frac{\overline{(\gamma-1)}}{[\overline{(\gamma-1)}]_{\rm nth}}
             = \!\! (\Gamma_{\rm inj}-1)(\Gamma_{\rm inj}-2)\!\!\!\!
               \int\limits_{\gamma_{\rm min}}^{\gamma_{\rm max}} 
               \!\!\!\! (\gamma - 1) \gamma^{-\Gamma_{\rm inj}} d \gamma,
\end{equation}
where the nominator corresponds to the energy given to accelerate the electrons between Lorentz factors $\gamma_{\rm min}$ and
$\gamma_{\rm max}$, and denominator corresponds to the energy given in the purely non-thermal injection.
For $\Gamma_{\rm inj}=3.0$ and $\gamma_{\rm min}=10$, the non-thermal fraction is $f_{\rm nth}=0.19$, and for 
$\gamma_{\rm min}=15$, $f_{\rm nth}=0.13$.
The rest of dissipated energy is given to electrons via stochastic heating as described in  sect.~4.1 of \citet{VP09}.
We also consider a purely thermal model, where 100 per cent of energy gained by electrons is due to stochastic heating.
For an illustration, we simulated an innermost zone ($i=1$) of the hot flow with parameters described in
Table~\ref{tab_params}.
The results are shown in Fig.~\ref{fig_nth_frac}.
We find that the resulting spectra  are very similar for  non-thermal  fractions $f_{\rm nth}>0.1$. 
For even smaller $f_{\rm nth}$, the spectra are harder, but slightly adjusting other parameters 
(magnetic field, Thomson optical depth) we can obtain spectra, which are very close to the original purely non-thermal models.
At the same time,  purely thermal model has a much harder X-ray spectrum (with spectral index $\alpha=-0.43$  
against $\alpha=-0.80$ for the non-thermal case) and higher equilibrium temperature 
($kT_{\rm e}=213$~keV against  $kT_{\rm e}=94$~keV for the non-thermal case), inconsistent with the existing X-ray data.

\begin{figure}
\epsfig{file=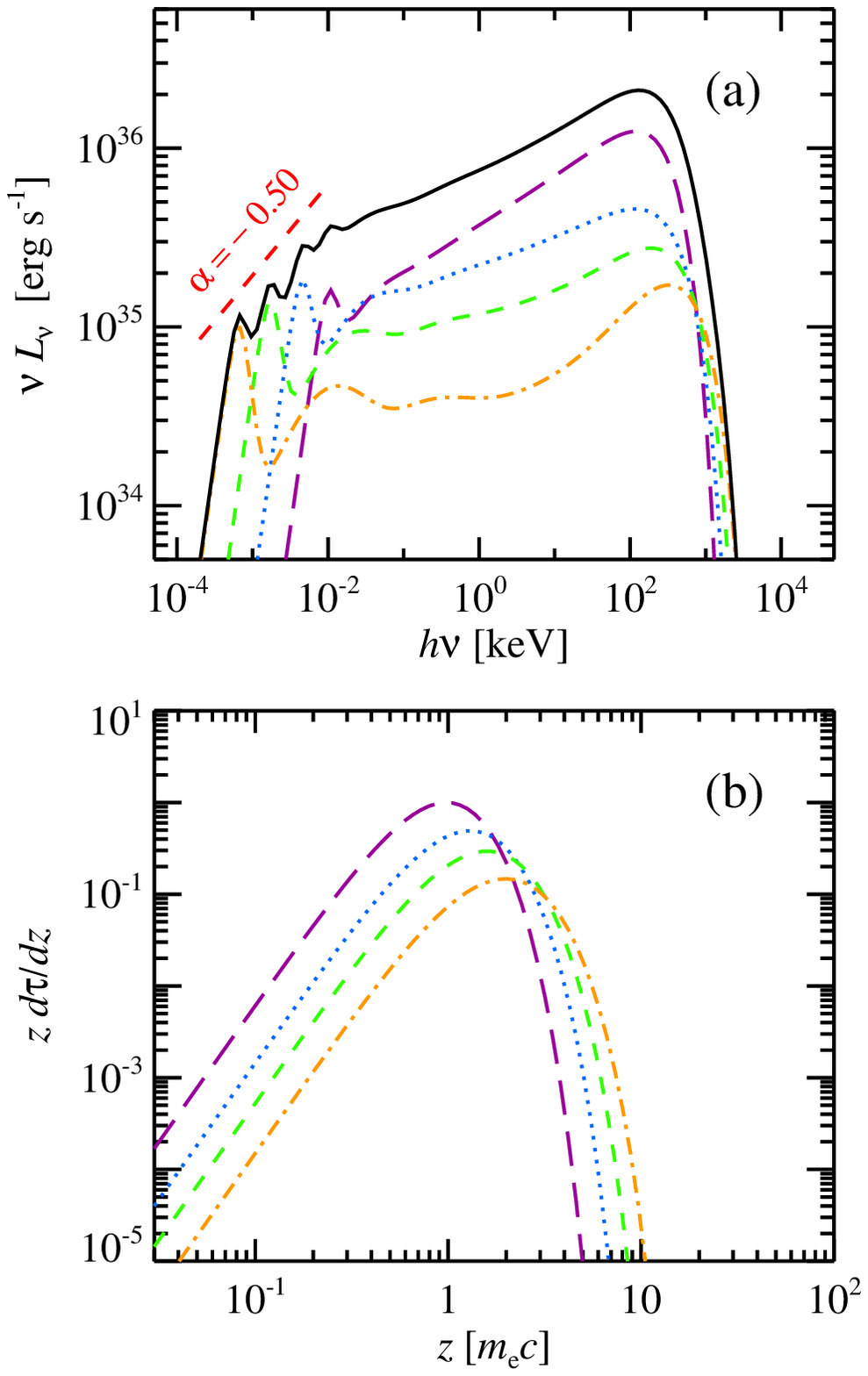, width=7cm}
\caption{
Same as in Figs~\ref{fig_pureSSC_inj2_5}~and~\ref{fig_pureSSC_inj3}, but for purely thermal electron distributions and 10 times
larger magnetic field in every zone.
The sharp peaks seen in the OIR are due to small number of zones considered and are not physical.
The OIR spectrum is likely to be smeared and represents a power-law at these wavelengths. 
The red dashed line is an analytical approximation for that.
}
\label{fig_th_OIR}
\end{figure}

In order to describe the data with purely thermal model, a much higher magnetic field is required.
Analysis of \citet{WZ00} showed that purely thermal SSC cannot be the dominant radiative mechanism in majority of the X-ray
binaries.
However, it is interesting to test how the OIR spectrum would change in this case.
Following equation 18 of \citet{WZ00}, the turnover frequency scales as
$\nu_{\rm t} \propto B^{0.91} \tau^{0.05} T_{\rm e}^{0.95}$.
As in the case of power-law electrons, it strongly depends on the magnetic field; however, the strongest dependence here is on
the electron temperature.
Assuming its scaling with radius $T_{\rm e}\propto R^{\kappa}$, we get
\begin{equation}
 \nu_{\rm t} \propto R^{0.95\kappa-0.91\beta-0.05\theta},
\end{equation}
which then implies that the OIR slope is 
\begin{equation}\label{eq:OIR_slope_th}
 \alpha_{\rm OIR, th} = \frac{4+4.75\kappa-3.55\beta-0.25\theta}{1.9\kappa-1.82\beta-0.1\theta}.
\end{equation}

As an illustration, we performed simulations for a purely thermal model for the whole inhomogeneous hot flow 
with the parameters from Table~\ref{tab_params}, except 
that the magnetic field in every zone was taken an order of magnitude larger, $B_{i,\rm th}=10 B_{i}$. 
This assumption is needed to match the typical X-ray spectra in the hard state ($\alpha\sim-0.7$).
However,  such high $B$ is above the upper limit determined by the equipartition with protons at a virial temperature. 
The equilibrium electron temperature determined by the heating=cooling condition is 
about 20 per cent larger than in the non-thermal case (with lower $B$). 
The spectra and the electron distributions  are shown in Fig.~\ref{fig_th_OIR}.
We find that the scaling of the electron temperature with radius can be approximated with $\kappa\sim0.3$.
Substituting $\beta=1.25$ and $\theta=0.5$, gives the resulting OIR slope $\alpha_{\rm OIR, th}\approx-0.5$, in good agreement
with the computed slope. Of course, no MeV emission is expected from purely thermal flows.

\subsection{Geometrical and radiative transfer effects}
\label{apndx:transfer}

\begin{figure}
\epsfig{file=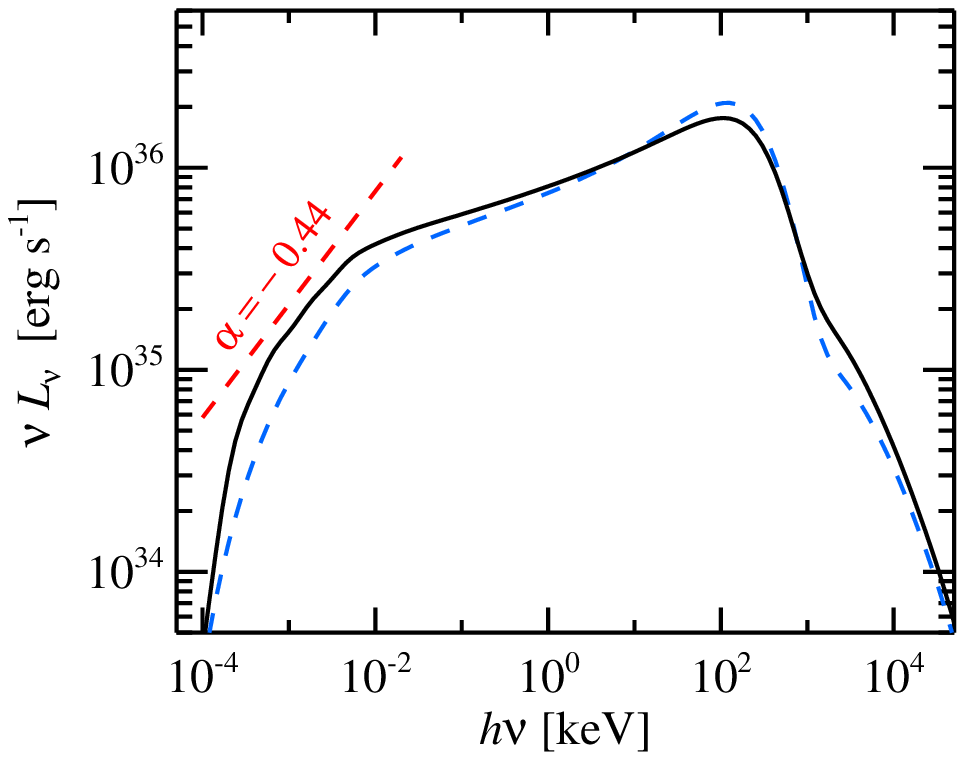, width=7cm}
\caption{
Spectra of a hot flow with different height-to-radius ratios.
Total spectrum for $H/R=0.25$ is shown with  blue dashed  line and for $H/R=0.5$ (same as in Fig.~\ref{fig_pureSSC_inj3}) is shown
with black solid  line.
The analytical approximation is shown with red dashed line.
}
\label{fig_HR}
\end{figure}

In this section, we investigate how the assumed geometry of the flow and 
the interaction between the zones affect the results. 
Let us first check how the thickness of the hot flow affects the spectra.
In order to approximate the hot flow with a number of tori, the condition of $\Delta R_{i}\approx2H_{i}$ should be satisfied.
Taking $H/R=0.25$ and using equation (\ref{eq:Rii1}) we find $R_{i+1}\approx5R_{i}/3$, with the innermost radius $R_1=3R_{\rm S}$. 
To cover the same distances from the BH, 
we now consider nine zones and scale parameters from Table~\ref{tab_params} so that they correspond to the distance 
to the centre of each zone. 
We compare the results of  calculations for  $H/R=0.25$  to those for $H/R=0.5$ in Fig.~\ref{fig_HR}.
Due to the increase of local electron number density ($H/R$ is reduced, while optical depth was kept constant), the X-ray
spectrum has become harder compared to the case $H/R=0.5$, while the OIR slopes are  almost identical.
The later is not surprising, as the OIR spectral shape is determined by scaling of the hot-flow parameters, 
rather than by their exact values in each zone.

\begin{figure}
\epsfig{file=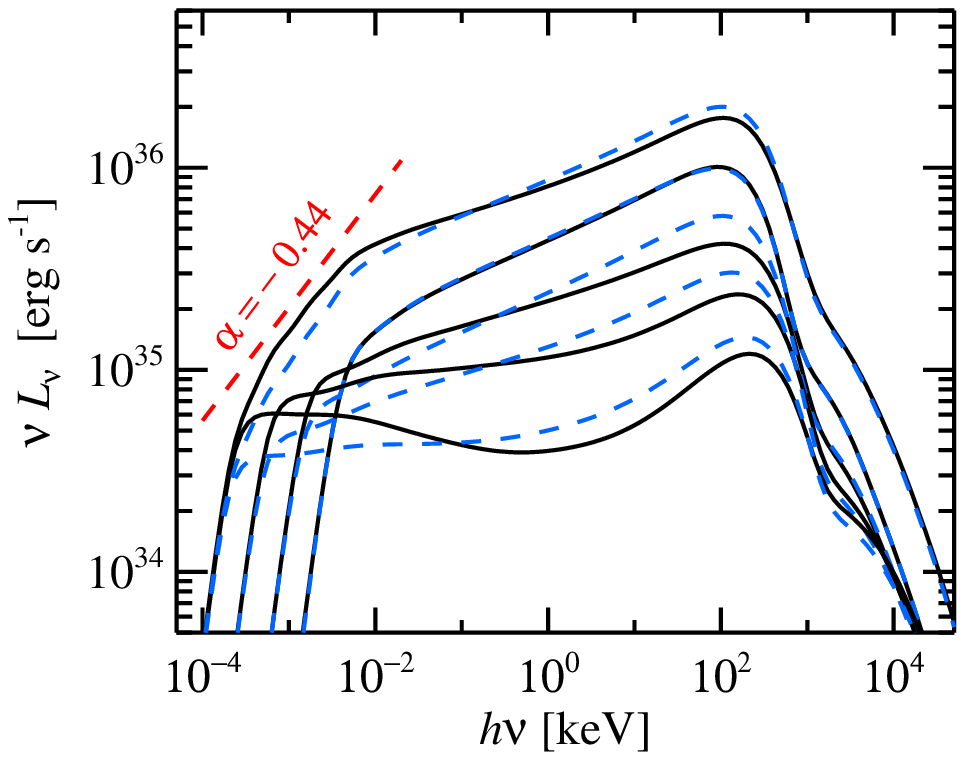, width=7cm}
\caption{
Spectra of the hot flow calculated accounting for the radiative transfer effects between the zones (blue dashed lines) 
and using local approximation (solid black lines). The upper lines represent the total spectra. 
Parameters of the zones are the same as in Fig.~\ref{fig_pureSSC_inj3} and $\Gamma_{\rm inj}=3.0$ is assumed.
The analytical approximation for OIR spectrum  (red dashed line) is sufficiently accurate for both cases. 
}
\label{fig_transfer}
\end{figure}

In our calculations we neglected the interaction between the zones and 
the spectrum of every zone was computed in the local approximation. 
However, photons from each zone can travel to the neighbouring zones. 
The largest  effect is expected for the outer zones, because they occupy a large solid angle 
as viewed from the inner zones and their luminosity is smaller than that of the  inner zones. 
The radiative transfer effects can be approximately accounted for by  
first computing spectra of each zone in local approximation and then by 
adding some fraction of the radiation escaping from each zones as an additional 
photon source for other zones. 
Luminosity coming from the $i$th zone to an adjacent $k=i\pm1$th zone is parametrized by 
$\displaystyle \Omega_{ik}/(4\pi) L_{i}$, where $\Omega_{ik}$ is the solid angle occupied by $k$th zone as seen from
$i$th zone. 
The formalism is similar to that used to simulate the outer cold disc photon injection to the hot flow.
Luminosity coming from the $i$th zone to a non-adjacent zone $j$ can be expressed as  
$\displaystyle \Omega_{ij}/(4\pi) L_{i} e^{-\tau_{\rm ij}}$, where $\tau_{ij}$ is the photon optical depth of the
medium between zones $i$ and $j$ (energy dependent, cumulative from different processes). 

The resulting spectra accounting for the radiative transfer effects are compared to those computed 
in local approximation  in Fig.~\ref{fig_transfer}. The changes in the OIR slope are hardly visible. 
The X-ray spectra of the outer zones became now harder and the variation of the 2--10 keV spectral index 
is  smaller $\Delta \alpha_{\rm X} \approx 0.03$ (vs. 0.08).
This effect is caused by additional photons coming from the harder-spectra inner zones and 
scattered in the outer zones in the direction to the observer. 
Thus,  the  spectra computed in the local approximation  are rather accurate and 
agree well with more detailed calculations.

\label{lastpage}

\end{document}